\documentclass{article}

\usepackage[english]{babel}

\usepackage[letterpaper,top=2cm,bottom=2cm,left=3cm,right=3cm,marginparwidth=1.75cm]{geometry}

\usepackage{wrapfig}
\usepackage{amsmath}
\usepackage{amssymb}
\usepackage{graphicx}
\usepackage{xcolor}
\definecolor{darkblue}{RGB}{0, 0, 128}
\usepackage[colorlinks=true, allcolors=darkblue]{hyperref}
\usepackage[authoryear]{natbib}
\usepackage{subcaption}

\newcommand{\diff}{\mathrm{d}}
\newcommand{\yobs}{y_{\text{obs}}}

\addto\extrasenglish{

}

\title{Simulations in Statistical Workflows}
\author{Paul-Christian Bürkner$^{1,*}$, Marvin Schmitt$^2$, Stefan T.~Radev$^3$}
\date{
$^1$ Department of Statistics, TU Dortmund University, Germany \\
$^2$ Independent Scientist \\
$^3$ Department of Cognitive Science, Rensselaer Polytechnic Institute, USA \\
$^*$ corresponding author, email: paul.buerkner@gmail.com
}

\begin{document}
\maketitle

\begin{abstract}
\noindent Simulations play important and diverse roles in statistical workflows, for example, in model specification, checking, validation, and even directly in model inference. Over the past decades, the application areas and overall potential of simulations in statistical workflows have expanded significantly, driven by the development of new simulation-based algorithms and exponentially increasing computational resources. In this paper, we examine past and current trends in the field and offer perspectives on how simulations may shape the future of statistical practice. \\ 

\end{abstract}

\section{Introduction}

Computer simulations have granted modern scientists inclusive access to ``would-be'' worlds created by devices of imaginary results \citep{good1950probability}. These worlds can be deterministic or random-like \citep{shannon1998introduction}; they can carry new insights or span an epistemic void \citep{grim2024comp}; they can merely augment or altogether redefine scientific models \citep{duran2020simulation}. Simulation can be modestly viewed as an aid to the golden path of experimentation \citep{guala2002models} or as a science in its own right -- \textit{a science of simulation} \citep{casti1996would}. Accounts of simulation as a ``new'' scientific method seem to emerge in regular intervals throughout the history of ideas, for instance, as part of the \textit{sciences of the artificial} \citep{simon1974sciences}, as part of a \textit{digital revolution in science} \citep{casti1996would}, or as part of \textit{simulation intelligence} encompassing an AI-permeated computational toolkit \citep{lavin2021simulation}.

In statistics, the emergence of simulation is typically associated with the birth of Monte Carlo methods during World War II \citep{robert2011short}, eventually leading to the publication of the Metropolis algorithm \citep{metropolis1953equation}.
\citet{hastings1970monte} and \citet{peskun1973optimum} later generalized the Metropolis algorithm as a family of simulation-based tools, namely, Markov chain Monte Carlo (MCMC), designed to mitigate the curse of dimensionality in direct Monte Carlo estimation.
Crucially, the inception of Monte Carlo methods coincides with the end of what \citet{efron2021computer} characterize as \textit{classical statistical inference} and the beginning of early computer-age methods, such as bootstrap and jackknife, that demand randomized resampling and mechanical repetition.

These early simulation methods greatly expanded the reach of inferential statistics, as they provided mechanical solutions to problems that were previously analytically intractable (e.g., high-dimensional expectations) or too cumbersome to execute manually (e.g., non-parametric estimation via bootstrap or permutation tests).
Moreover, it was not until general-purpose computers had shrunk considerably in size that Bayesian inference became more than ``...a macho activity enjoyed by those who were fluent in definite integration'' \citep[][p.~319]{mackay2003information}.
In a way, digital simulation fulfilled Pearson's dream of a universal tool for carrying out statistical experiments with synthetic coins and roulettes \citep{kucharski2016perfect}, ushering in the widespread application of both frequentist and Bayesian paradigms.

However, we argue that Bayesian methods have gained the most from the transition to computational statistics and the rise of simulation-based tools. Unlike frequentist approaches, which often rely on asymptotic approximations, Bayesian methods require the evaluation of high-dimensional posterior distributions that are rarely analytically tractable \citep{gelman_bayesian_2013}.
Put simply, Bayesian inference is hard because integration is hard.
Thus, the same MCMC methods developed for approximating energy distributions in statistical physics have proven instrumental for sampling from \textit{conditional distributions} arising in Bayesian analysis \citep{robert2011short}.

Early MCMC methods freed Bayesian analysis from the confines of conjugate models and closed-form posteriors, most notably through the introduction of the Gibbs sampler \citep{geman1984stochastic, gelfand1990sampling} and hybrid variants \citep{tierney1994markov}. However, these methods still required \textit{explicit models} with tractable likelihoods. In contrast, \textit{implicit models}---defined through a stochastic mechanism (e.g., a randomized simulation) without a known distribution for computing the likelihood of their outputs \citep{diggle1984monte}---remained out of reach for Bayesian methods until the advent of approximate Bayesian computation \citep[ABC;][]{rubin1984bayesianly, tavare1997inferring}.
In ABC, stochastic simulation explicitly takes on a dual role: it serves both as an epistemic tool for solving \textit{forward problems} (from unknowns to knowns) and as a computational tool for solving \textit{inverse problems} (from knowns to unknowns).
Indeed, the coinage of the term simulation-based inference \citep[SBI;][]{cranmer2020frontier} is reflective of this role.

\begin{wrapfigure}[13]{r}{0.4\textwidth}
  \vspace*{-1.3\baselineskip}
  \begin{center}
    \includegraphics[trim={0 11.5cm 18cm 0},clip, width=1.0\linewidth]{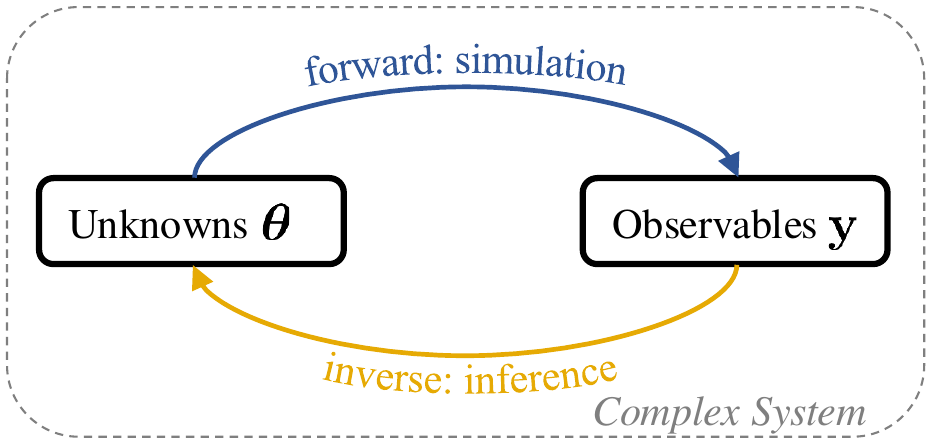}
  \end{center}
  \vspace*{-1.0\baselineskip}
  \caption{Forward simulation and inverse inference are two central components in a statistical model of unknowns~$\theta$ and observables~$y$.}
\end{wrapfigure}
As the notion of statistical model loses its well-defined contours in an AI-dominated conceptual landscape, the role of simulation in statistics becomes increasingly difficult to delineate from its role in science in general. However, since this paper focuses on statistical workflows—iterative processes of model building and model criticism—our discussion will center on the versatility of simulating \textit{from} statistical models. In particular, this means generating random draws from model-implied distributions, much like drawing random cards from a cleverly arranged deck. Given this focus, we will not detail sampling algorithms that primarily rely on evaluating the \textit{density} of statistical models rather than their generative capabilities.

Among density-based approaches, the most prominent family of algorithms is undoubtedly MCMC \citep{brooks2011handbook}. MCMC has seen rapid developments in recent years, including locally adaptive methods that can handle distributions with complex geometries \citep{biron2024automala, modi2024delayed} and methods that leverage modern hardware for large-scale parallelization \citep{margossian2024nested, sountsov2024running}. Other popular families of algorithms are Sequential Monte Carlo \citep[SMC;][]{del2006sequential} and importance sampling \citep{tokdar2010importance}.
Even though these algorithms involve simulation steps, they typically do not require direct draws from the model’s generative distribution, but instead rely on simulating proposals or particles guided by density evaluations.

The organization of the following sections elucidates use cases for statistical model simulations throughout the key stages of a minimalist statistical workflow, namely: 1) model specification; 2) model verification; 3) model inference; and 4) model checking. We conclude with a brief outlook on simulation intelligence and the ascent of automated statistics.


\section{Model Specification}\label{sec:model-specification}

In a previous work, we argued that Bayesian modeling has evolved beyond traditional likelihood-prior formulations to encompass complex simulation-based inference techniques, posterior approximators, and amortized learning strategies \citep{burkner_models_2023}. To structure this expanding landscape, we proposed the PAD taxonomy, which categorizes Bayesian models along three fundamental axes:
\begin{itemize}
    \item P (\textbf{P}robability Distribution) -- the probabilistic joint model (hereafter called P model) that defines the relationships between model parameters, latent variables, and observed data.
    \item A (Posterior \textbf{A}pproximator) -- the inference algorithm used to approximate the posterior distribution, ranging from MCMC methods to neural networks.
    \item D (Training \textbf{D}ata) -- the observed data used to condition the model and update beliefs.
\end{itemize}
According to the PAD model taxonomy, \textit{model specification} refers to defining the components of a P model. In a Bayesian setting, this entails formulating a prior $\pi(\theta)$ and an observation model $\pi(y \mid \theta)$ which together determine the generative joint model $\pi(\theta, y)$. In a frequentist setting, the P model reduces to the observation model $\pi(y \mid \theta)$, which is defined over a specific domain $\Theta \ni \theta$.

As part of a Bayesian workflow \citep{gelman2020bayesian}, the plausibility of P models can already be visualized and evaluated through prior predictive or prior pushforward checks \citep{gabry2019visualization}. The purpose of these checks is to assess whether the generative behavior of a P model is consistent with the available domain expertise via simulations from the model-implied (i.e., prior predictive) distribution, $\pi(y) = \int \pi(y \mid \theta) \pi(\theta) \diff \theta$.
Consistency with domain expertise (i.e., known knowns and known unknowns) not only figures in domain-specific Bayesian workflows \citep{schad2021toward}, but can also be considered as a crucial aspect of iterative theory building in general.

Here, once again, simulation serves as an antidote to complexity: consistency with domain expertise may not be immediately evident from the model formulation alone, even when a model's specification appears beguilingly simple \citep[see, for example,][for a probabilistic extension of Conway's Game of Life]{aguilera2019probabilistic}.
Moreover, if the data $y$ is high-dimensional, the utility of consistency checks can be challenging to assess.
This is where \textit{prior pushforward checks} come into play \citep{gabry2019visualization}, following the procedure suggested by \citet{schad2021toward}: (1) define a low-dimensional, interpretable summary statistic $T(y)$; (2) determine plausible value regions for the summary statistic; and (3) simulate model predictions to ensure the summary statistic falls within these regions\footnote{In a frequentist setting, the prior can be replaced with a set of plausible values for the parameters, $\Theta_{\text{plausible}}$.}.

\begin{figure*}[t]
    \centering
    \includegraphics[width=.99\linewidth]{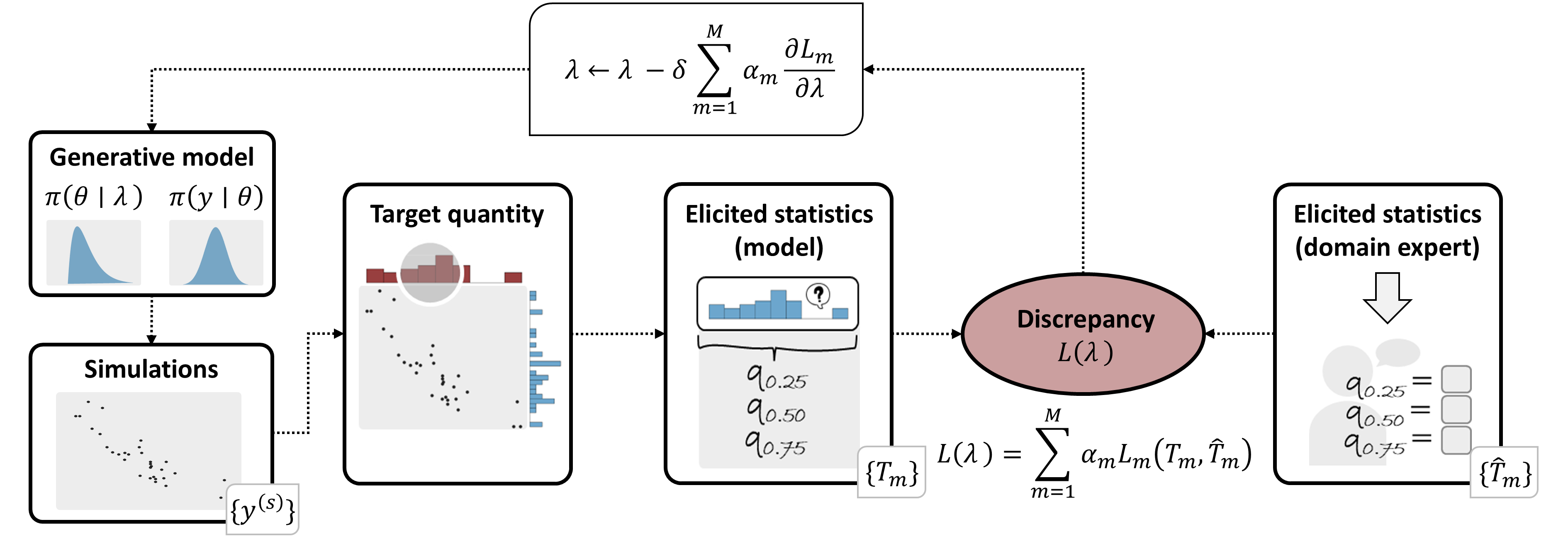}
    \caption{Graphical illustration of the simulation-based framework for prior elicitation proposed by \citet{bockting2024simulation}. The process starts by identifying target quantities via the domain expert and eliciting expert statistics $\hat{T}$. Model predictions are then simulated by sampling from a parametric prior $\pi(\theta \mid \lambda)$ and computing model-implied target quantities $T$. A loss function $L$ assesses the consistency between model and expert-elicited statistics and adjusts prior hyperparameters $\lambda$ to minimize discrepancies. The process continues until prior predictions align closely with expert knowledge.}
    \label{fig:prior_elicitation}
\end{figure*}

Finally, the simulation-based procedure outlined above can be inverted for the purpose of \textit{prior elicitation}, which seeks solutions to the hard problem of transforming non-probabilistic domain knowledge into well-defined prior distributions \citep{mikkola2024prior}.
Accordingly, recent optimization-based approaches attempt to \textit{learn} the consistency between expert knowledge and model predictions by minimizing discrepancies between expert-elicited and model-generated summary statistics \citep[][see also \autoref{fig:prior_elicitation}]{bockting2024simulation, hartmann2020flexible}.
Most recently, generative networks have been employed to infer non-parametric joint priors from sparse expert knowledge and synthetic data \citep{mikkola2024preferential, bockting2024expert}, opening new avenues for creative utilization of model simulations.

\section{Model Verification}\label{sec:model-verification}

Before deploying a statistical model on real data, we should first ensure that the interplay between the probabilistic joint model P and the approximator A functions as intended in the scenarios they were designed for, namely, when their underlying assumptions hold. In many cases, simulations offer the only feasible means of verifying such correctness, or at least gauging \textit{in silico} whether inferences are sufficiently informative for their intended purposes. 

\begin{figure}[t]
    \centering
    \includegraphics[width = 0.90\textwidth]{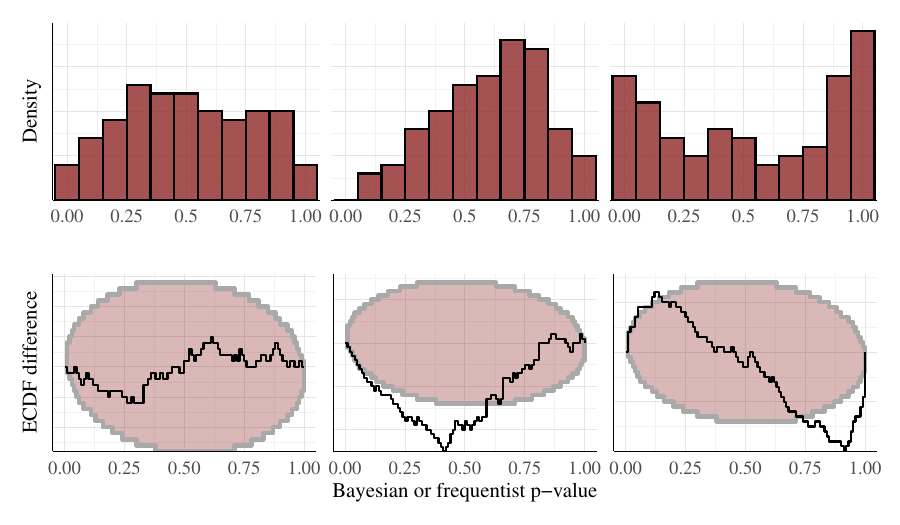}
    \caption{Simulation-based $p$-value histograms (top) and corresponding empirical cumulative distribution function (ECDF) difference plots \citep{sailynoja_graphical_2022} (bottom) for three hypothetical quantities of interest. The pink areas in the ECDF difference plots indicate 95\%-confidence intervals under the assumptions of uniformity and thus allow for a null-hypothesis significance test of Bayesian and frequentist model calibration. Left: A well-calibrated quantity. Center: A miscalibrated quantity with too few small $p$-values. When testing for self-consistency, this indicates a positive bias in the quantity's estimates. Right: A miscalibrated quantity with too many extreme $p$-values. When testing for self-consistency, this indicates overconfident uncertainty estimates (i.e., underdispersion).}
    \label{fig:SBC}
\end{figure}

\subsection{Verifying Calibration}
\label{sec:model-verification:calibration}
Proper uncertainty calibration is is crucial for ensuring confidence in estimates and predictions from statistical models \citep{little_calibrated_2006, gneiting_probabilistic_2007}. Informally, we can say that the uncertainty of an approximator is well calibrated if it captures the true amount of uncertainty in a system. Except for special cases, accessing this true uncertainty is only ever possible with simulations. Below, we formalize the verification process for uncertainty calibration for both Bayesian and frequentist perspectives.

\subsubsection{Bayesian Calibration}
\label{sec:model-verification:sbc}

In the following, we will denote an approximator of the true posterior $\pi(\theta \mid y)$ as $q(\theta \mid y)$. This $q(\theta \mid y)$ may be given implicitly, for example, in the form of an MCMC algorithm or explicitly, for example, in the form of a generative neural network.
The accuracy of an approximator $q(\theta \mid y)$ is difficult to study analytically, since the true posterior $\pi(\theta \mid y)$ is rarely known in the first place. Fortunately, certain Bayesian self-consistency properties can be used even without knowledge about $\pi(\theta \mid y)$. Consider, for instance, a target quantity of interest $T = T(\theta)$, which may be derived from the parameters $\theta$. This could include the parameters themselves or any pushforward quantity, such as posterior predictions. For all target quantities $T$ and all uncertainty regions $U_\alpha(T(\theta) \mid y)$ obtained from $\pi(\theta \mid y)$ with nominal coverage probability $\alpha$ (e.g., credible intervals based on posterior quantiles), we know that
\begin{equation}
\label{prior-cond-sbc}
  \alpha = \int \int \mathbb{I}(T^* \in U_\alpha(T \mid y)) \, \pi(y \mid \theta^*) \, \pi(\theta^*) \, \diff y \, \diff \theta^*,
\end{equation}
where $\theta^* \sim \pi(\theta^*)$ are prior draws that serve as the ground truth for the corresponding target quantity $T^* = T(\theta^*)$.
In other words, the probability that an uncertainty region with coverage probability $\alpha$ contains the true value must be equal to $\alpha$ on average over the data-generating process. For a univariate quantity $T$, we can formulate this requirement as follows: under perfect Bayesian calibration, the posterior probability $p := \Pr_{\pi(\theta \mid y)}(T \leq T^*)$ (i.e., the ``Bayesian $p$-value'') is uniformly distributed between 0 and 1 \citep{cook2006validation, gneiting_probabilistic_2007, talts_validating_2018}.
For a good approximation $q(\theta \mid y)$ of the true posterior $\pi(\theta \mid y)$, the above property should hold within tolerable error bounds.
Unfortunately, the associated integrals in \autoref{prior-cond-sbc} are not analytic, so none of their properties can be verified directly for $q(\theta \mid y)$.
However, we can employ simulations from the underlying P model. For a data set $y$, we define
\begin{equation}
   p(T^*, T^{(1:M)} \mid y) := \frac{1}{M} \sum_{m=1}^M \mathbb{I}(T^{(m)} \leq T^*),
\end{equation}
where $\theta^{(m)} \sim q(\theta \mid y)$ are $M$ draws from the approximate posterior and $T^{(m)} = T(\theta^{(m)})$.
If the posterior approximator is equal to the true posterior, the distribution of the posterior probabilities $p(T^*, T^{(1:M)} \mid y)$ will follow a discrete uniform distribution between 0 and 1 \citep{modrak2025simulation}.

To test this property empirically, we first sample $S$ parameter draws from the prior, ${\theta^*}^{(s)} \sim \pi(\theta^*)$ and subsequently $S$ data sets $y^{(s)} \sim \pi(y \mid {\theta^*}^{(s)})$ from the likelihood. Then, we obtain $M$ draws $\theta^{(m)} \sim q(\theta \mid y^{(s)})$ from the approximate posterior to compute $p^{(s)} := p(T^*, T^{(1:M)} \mid y^{(s)})$. Finally, we can test the set of posterior probabilities $\{ p^{(s)} \}$ for uniformity (see \autoref{fig:SBC} for an illustration).
More powerful methods for diagnosing \textit{joint calibration} are possible, including training a classifier to learn test statistics directly from simulations \citep{yao2023discriminative, bansal2025surprising} or using the model likelihood if available \citep{modrak2025simulation}.

These procedures for checking Bayesian calibration are collectively referred to as simulation-based calibration (SBC) \citep{talts_validating_2018, modrak2025simulation}. Crucially, SBC requires \textit{nested simulations}: in the outer loop, we simulate parameters ${\theta^*}^{(s)}$ and data $y^{(s)}$ from the joint model $\pi(\theta^*) \, \pi(y \mid \theta^*)$; in the inner loop, we sample $\theta^{(m)}$ from the approximate posterior $q(\theta \mid y^{(s)})$. As a result, SBC is typically computationally intensive, yet it remains a general method for validating approximate posteriors.

\subsubsection{Frequentist Calibration}

Frequentist calibration works very similarly to Bayesian calibration, with two main differences. First, we do not sample true parameter values from a prior, $\theta^* \sim \pi(\theta^*)$, but rather treat $\theta^*$ as fixed at some value. Second, our target of inference is not the true posterior $\pi(\theta \mid y)$, but a point estimator $\hat{\theta} = \hat{\theta}(y)$ of the parameters, along with the true sampling distribution $\pi(\hat{\theta} \mid \theta^*)$ of the point estimator. This, in turn, implies a point estimator $\hat{T} = T(\hat{\theta})$ for any quantity of interest $T$, with a corresponding true sampling distribution $\pi(\hat{T} \mid T^*)$. The latter can be transformed into uncertainty regions $U_\alpha(\hat{T} \mid y)$ (i.e., confidence intervals), which, under perfect frequentist calibration, will contain the true value $T^*$ for $\alpha$ percent of the datasets generated from $\pi(y \mid \theta^*)$:
\begin{equation}
\label{prior-cond-freq}
  \alpha = \int \mathbb{I}(T^* \in U_\alpha(\hat{T} \mid y)) \, \pi(y \mid \theta^*) \, \diff y.
\end{equation}
If $T$ is univariate, we again obtain an equivalent but simpler formulation: under perfect frequentist calibration, the sampling distribution probability $p := \mathbb{P}_{\pi(\hat{T} \mid T^*)}(\hat{T} \leq T^*)$ (i.e., the $p$-value) is uniformly distributed between 0 and 1 \citep{rosenblatt_remarks_1952, gneiting_probabilistic_2007}. This enables testing the calibration of an approximator $q(\hat{T} \mid T^*)$ of the true sampling distribution $\pi(\hat{T} \mid T^*)$ via simulations: sample $S$ draws $y^{(s)} \sim \pi(y \mid \theta^*)$, compute $p^{(s)} = \mathbb{P}_{q(\hat{T} \mid T^*)}(\hat{T} \leq T^*)$ using the approximate sampling distribution, and then test the resulting set of $p$-values $\{ p^{(s)} \}$ for uniformity.

\subsubsection{Power Analysis}

The frequentist self-consistency of $p$-values holds only if the true parameter value $\theta^*$ used for simulating data from the likelihood $\pi(y \mid \theta^*)$ is the same parameter assumed in the sampling distribution $\pi(\hat{\theta} \mid \theta^*)$. In a null-hypothesis significance framework (see also \autoref{hypothesis-testing}), we consider a sampling distribution $\pi(\hat{\theta} \mid \theta_0)$ given the parameter value $\theta_0$ characterizing the null-hypothesis. If the $\theta^*$ assumed in the sampling process $\pi(y \mid \theta^*)$ is different from $\theta_0$, uniformity of the $p$-value distribution is no longer expected. Instead, when $\theta^*$ represents an explicit alternative hypothesis, the $p$-value distribution measures statistical power. Simulations are the prime tool for performing power analysis since analytical expressions of sampling distribution are often only available for few special parameter values, usually representing the null-hypothesis \citep{cohen_statistical_2013}. Simulation-based power analysis can also be performed in a Bayesian framework by choosing a prior $\pi(\theta_0)$ within the estimated model that differs from the prior $\pi(\theta^*)$ used in the data-generating process.

\subsection{Other Simulation Studies}

Good uncertainty calibration is undoubtedly an important property that should ideally be verified before entrusting the pair of probabilistic model and approximator with performing inference on real data. However, calibration alone is often insufficient. For example, when only testing parameters for calibration in a Bayesian model, $T(\theta) = \theta$, then not only the posterior $\pi(\theta \mid y)$ but also the prior $\pi(\theta)$ would pass the above described uniformity tests \citep{modrak2025simulation}. Accordingly, we have to additionally study the \textit{sharpness} of uncertainty estimators \citep{gneiting_probabilistic_2007}, which essentially indicates how small the uncertainty regions are for any fixed $\alpha$ \citep[see also][]{burkner_models_2023}. Among all well-calibrated uncertainty estimators, we would then choose the one that is sharpest. As for calibration, sharpness is almost always non-analytic so simulations are the only way to study it.

Even when not considering uncertainty at all, but just studying the accuracy of point estimators, simulations are required. We can measure the accuracy of a point estimator $\hat{T}$ as the distance $D$ to its target $T^*$ averaged across the data-generating process:
\begin{equation}
\bar{D}(\hat{T}, T^*) := \int D(\hat{T}(y), T^*) \, \pi(y \mid \theta^*) \, \diff y
\end{equation}
For example, when $D$ is the squared difference, this leads to the mean squared error (MSE):
\begin{equation}
    \text{MSE}(\hat{T}, T^*) := \int (\hat{T}(y) - T^*)^2 \, \pi(y \mid \theta^*) \, \diff y \approx \frac{1}{S} \sum_{s=1}^S (\hat{T}(y^{(s)}) - T^*)^2,
\end{equation}
approximated via draws $y^{(s)} \sim \pi(y \mid \theta^*)$. The same can be done for a Bayesian point estimator (e.g., the posterior mean or median) by first sampling ${\theta^*}^{(s)} \sim \pi(\theta^*)$ and then $y^{(s)} \sim \pi(y \mid {\theta^*}^{(s)})$.

\section{Model Inference}\label{sec:model-inference}

Simulations can also be used directly for model inference \citep{cranmer2020frontier}, that is, to estimate parameters or obtain decisions from the triple of statistical model, approximator, and observed data.

\subsection{Hypothesis testing}
\label{hypothesis-testing}

\begin{figure}[tp]
    \centering
    \includegraphics[width = 0.94\textwidth]{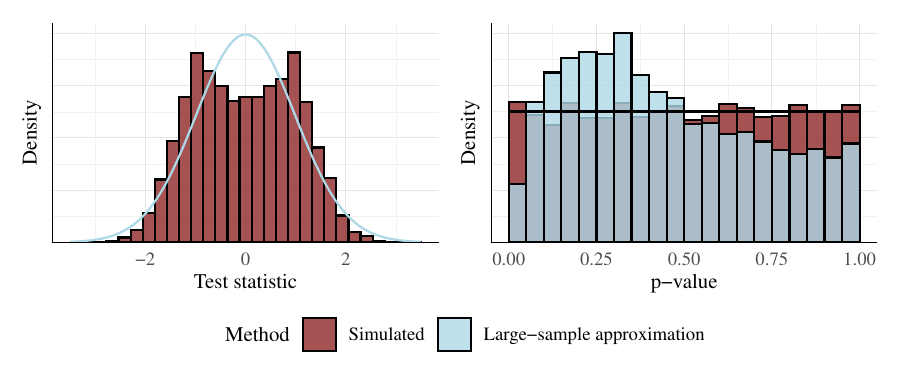}
    \caption{Illustration of a simulation-based test. In the chosen scenario, the means of two groups are compared using a two-sample $t$-test for equal variances. Two independent data sets, each consisting of $N=40$ observations, were simulated from $\text{LogNormal}(\mu = 2, \sigma = 2)$. The null hypothesis asserts equal group means and equal variances. However, the data deviate substantially from normality. Accordingly, even for $N=40$ per group, the left plot shows that the true distribution of the test statistics (approximated via simulations; red histogram), is clearly different from the large sample approximation assumed by the t-test (blue density). The right plot shows that the corresponding $p$-value distribution under the null hypothesis is highly non-uniform for the large sample approximation (blue histogram). In contrast, $p$-values obtained via a simulation-based test ($S=10,000$) are almost perfectly uniform, with only small random error (red histogram).}
    \label{fig:sim-test}
\end{figure}

Every statistical test is based on a test statistic $T = T(y)$ that extracts the specific property of interest from the data $y$. For example, if the goal is to compare the location of two groups, then $T(y)$ could be the difference between the two group means. In addition to the test statistic itself, its distribution $\pi(T(y) \mid \theta)$ implied by the likelihood $\pi(y \mid \theta)$ needs to be analytically tractable, at least for some reference parameter value $\theta_0$ that captures the null hypothesis (e.g., no difference). 

In some cases, the sampling distribution under the null hypothesis is indeed analytic, for instance, as in $t$-tests for normally distributed data, but often enough it is not.
We may then appeal to asymptotic approximations, under which the (sampling) distribution of the test statistic is a well-characterized distribution in the limit of an infinite number of simulations \citep{ibragimov_statistical_2013}.
However, the accuracy of these approximations for smaller sample sizes can be questionable (see \autoref{fig:sim-test}).

Simulation-based tests provide a powerful alternative to analytic solutions or large-sample approximations \citep{racine_simulation-based_2007, vasishth_foundations_2010}. They generally proceed as follows: Sample $S$ draws $y_0^{(s)} \sim \pi(y \mid \theta_0)$ from the likelihood given $\theta_0$. Then, compute the test statistic $T_0^{(s)} := T(y_0^{(s)})$ for each draw, which implies $T_0^{(s)} \sim \pi(T(y) \mid \theta_0)$. Thereby, we gain access to samples from the target distribution $\pi(T(y) \mid \theta_0)$ simply via predefined transformations of individual samples (see \autoref{fig:sim-test} left).
Then, we compare the test statistic value $\tilde{T} := T(\yobs)$ from the observed data $\yobs$ with the distribution of simulated test statistic values $\{ T_0^{(s)} \}$.

Suppose we are interested in a one-sided test for lower values of $T$.
We would first compute $\tilde{r} = \sum_{s=1}^S \mathbb{I}( T_0^{(s)} < \tilde{T})$, where $\mathbb{I}$ is the indicator function. Then, we would compute the $p$-value $\tilde{p}$ corresponding to $\yobs$ as the normalized rank $\tilde{p} := \tilde{r} /S$ (see \autoref{fig:sim-test}, right) and the significance threshold $T_{\rm \alpha}$ as the empirical $\alpha$-percentile of a set $\{ T_0^{(s)} \}$ of simulated test statistics.

This approach is very general and can even be applied on top of Bayesian methods \citep{schmitt_meta-uncertainty_2023}. However, it relies on two important prerequisites. First, we need to be able to sample efficiently from the likelihood, so that we can choose $S$ large enough to obtain sufficiently accurate approximations of the $p$-value. Second—and often more prohibitive—we require a suitable test statistic $T(y)$ that captures the effect of interest independently of potential nuisance parameters and is straightforward to compute directly from the data. If $T$ is a function of model parameter estimates, that is, $T = T(\hat{\theta}(y))$, then performing simulation-based testing becomes computationally intensive: obtaining model estimates for a single dataset is often already slow, and doing so $S$ times may render the entire procedure infeasible. Taken together, these challenges have thus far limited the widespread adoption of simulation-based testing. Fortunately, efficiency issues can be addressed by \textit{amortized methods}, which we discuss in \autoref{sec:parameter-estimation} as a special case of simulation-based inference.

\begin{figure*}[t]
    \centering
    \includegraphics[width=.99\linewidth]{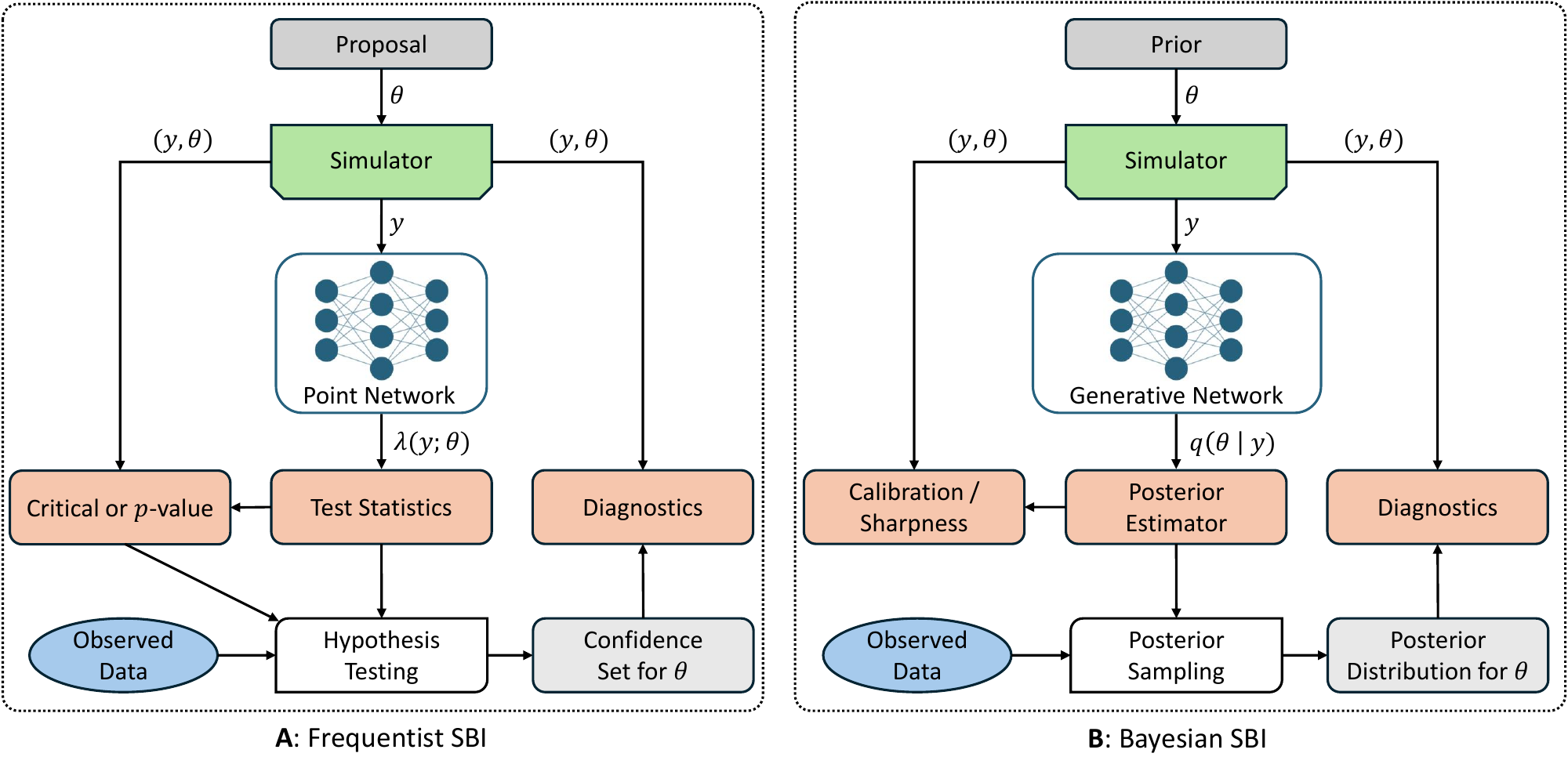}
    \caption{Graphical illustration of frequentist vs. Bayesian approaches to simulation-based inference (SBI). \textbf{Panel A}: In frequentist SBI, a classifier is trained on simulated data and parameters $(y, \theta)$ to efficiently estimate a test statistic $\lambda(y;\theta)$ (e.g., the odds function). This statistic is used to compute critical and \textit{p}-values via further simulations. An approximate confidence set for $\theta$ is constructed by inverting a series of hypothesis tests. The coverage of this set can be diagnosed through additional simulations \citep[adapter after][]{dalmasso2024likelihood}. \textbf{Panel B}: In Bayesian SBI, a generative neural network is trained to estimate the posterior $\pi(\theta \mid y)$ from simulated $(y, \theta)$ pairs. The calibration and sharpness of the resulting estimator can be evaluated via further simulations. Given observed data, the estimator can rapidly sample from the approximate posterior. Further simulations can be used to evaluate the fidelity of these samples via predictive checks or out-of-distribution (OOD) detection.}
    \label{fig:sbi_parameter_estimation}
\end{figure*}

\subsection{Parameter Estimation}
\label{sec:parameter-estimation}

In parameter estimation, we construct estimators that approximate the unknown parameters of a P model using the data in an attempt to solve the underlying inverse problem (i.e., estimating the unknowns from the knowns).
There are myriad ways to construct useful estimators. Two popular approaches rooted in the frequentist and Bayesian frameworks, respectively are: 1) constructing well-calibrated \textit{confidence sets} and 2) recovering the \textit{posterior distribution} by updating the prior $\pi(\theta)$ with data. For the sake of clarity, we briefly reiterate the goals of these approaches.
\vspace{0.5em}

\noindent \textit{Confidence sets}: For any parameter vector $\theta$ and likelihood function $\pi(y \mid \theta)$, we are interested in constructing a random confidence set $\mathcal{R}(y)$ with nominal $1 - \alpha$ coverage, such that
\begin{equation}
    \mathbb{P}_{\pi(y \mid \theta)}(\theta \in \mathcal{R}(y)) \geq 1 - \alpha \quad \forall \theta \in \Theta.
\end{equation}

\noindent \textit{Posterior distributions}: For a given prior $\pi(\theta)$, we are interested in updating the prior to the posterior $\pi(\theta \mid y)$, which incorporates all information about $\theta$ carried by the data $y$,
\begin{equation}
    \pi(\theta \mid y) = \pi(y \mid \theta) \, \pi(\theta) \, \pi(y)^{-1}.
\end{equation}

Needless to say, both problems are computationally challenging for non-analytic models (e.g., no closed-form for the Bayesian posterior or unknown frequentist sampling distribution).
The situation is further aggravated by P models defined \textit{solely} through a (randomized) algorithm for generating data $y$ from parameter configurations $\theta$ instead of explicitly assuming a parametric data model $\pi(y \mid \theta)$ \citep{diggle1984monte}.
For such \textit{implicit models}, the construction of confidence sets is challenging not only because we cannot evaluate the likelihood, but also because we need to test null hypotheses across the entire parameter space \citep{dalmasso2024likelihood}. By the same token, posterior approximation is difficult because not only the marginal likelihood $\pi(y)$ involves an intractable integral, but also the implicit likelihood \citep{cranmer2020frontier}, rendering posterior approximation doubly intractable.

\paragraph{Approximate Bayesian Computation} 
Model simulations can be used to connect even intractable models to real data while remaining true to the likelihood principle \citep{diggle1984monte, cranmer2020frontier}.
Arguably, the most popular approach to simulation-based inference (SBI) is approximate Bayesian computation \citep[ABC;][]{marin2012approximate}. 
The original ABC rejection sampler approximates the posterior by repeatedly proposing parameters from the prior $\pi(\theta)$ and then simulating a data set from the data model $\pi(y \mid \theta)$. If the resulting data set is sufficiently similar to the actually observed data set $\yobs$, the corresponding parameter vector is retained as a sample from the posterior, otherwise rejected. More sophisticated versions include likelihood-free MCMC \citep[ABC-MCMC;][]{marjoram2003markov, picchini2014inference}, Sequential Monte Carlo \citep[ABC-SMC;][]{beaumont2009adaptive, del2012adaptive}, and various hybrid methods \citep{picchini2024guided}.

Despite their theoretical elegance \citep{frazier2018asymptotic}, most ABC methods share the fundamental limitation of all non-amortized methods: computations must be repeated from scratch each time a model is fit to new data. Yet, there are many scenarios in which multiple model refits are necessary—for example, when modeling multiple data sets \citep{von2022mental}, performing \textit{in silico} model verification (\autoref{sec:model-verification}), or aiming for real-time inference \citep{zeng2025real}.
To address these challenges, a different way of utilizing simulations was needed---a way to pool or ``compile'' computations into global estimators that can produce near-instant results for arbitrary queries \citep{gershman2014amortized, paige2016inference, le2017inference}. In hindsight, neural networks, as modular and flexibly composable universal function approximators, proved to be the ideal choice. 

\paragraph{Amortized Inference via Neural Networks}

Amortized inference asks how to flexibly reuse inferences or estimators in order to answer numerous queries without recomputation overhead.
It has been proposed as a model of human probabilistic reasoning \citep{gershman2014amortized}, a method for fast inversion of graphical models \citep{stuhlmuller2013learning}, and a means of learning variational posteriors with neural networks \citep{kingma2013auto, rezende2014stochastic}.
For statistical inference with ``white-box'' models, the latter idea extends to using simulations as training data for neural networks that approximate quantities of interest, such as point estimates, test statistics, or full posterior distributions \citep[][see also \autoref{fig:sbi_parameter_estimation}]{burkner_models_2023, zammit2024neural}. Neural networks are particularly favorable for amortized inference if the statistical model is sufficiently high dimensional (in data and/or parameter space), no good hand-crafted summary statistics are available, and model simulations can be obtained in reasonable time (to provide sufficient training budget).

In a frequentist setting, simulation-based training can be employed to amortize the computation of arbitrary test statistics $\lambda(y; \theta)$ \citep[][see also left panel of \autoref{fig:sbi_parameter_estimation}]{dalmasso2024likelihood}. For instance, one viable approach is to re-frame odds ratio estimation as a classification problem, in which case we can train a neural classifier to approximate the likelihood ratio \citep{hermans2020likelihood} or odds function \citep{dalmasso2024likelihood} based on simulations.
Subsequently, we can learn a quantile regression of $\lambda(y; \theta)$ on $\theta$ that can approximate the critical value $C_\alpha$ for every $\alpha$-level, allowing us to construct approximate confidence sets of the form
\begin{equation}
    \widehat{\mathcal{R}}(\yobs) := \{\theta \in \Theta \mid \lambda(\yobs; \theta) \geq \widehat{C}_{\alpha}\}.
\end{equation}
\noindent Finally, the empirical coverage of frequentist confidence sets for every $\yobs$ can be assessed on a further held-out set of simulations \citep{dalmasso2024likelihood}.

In amortized Bayesian inference (ABI), a generative network seeks to learn a global posterior functional $q(\theta \mid y)$ for any observation $y$. Typically, the network would minimize a strictly proper scoring rule $\mathcal{S}$ in expectation over the joint distribution $\pi(\theta, y)$ of the P model:

\begin{equation}\label{eq:expected_score}
    \mathcal{L}(q) := \mathbb{E}_{\pi(\theta, y)}\Big[\mathcal{S}(q(\cdot \mid y), \theta)\Big] \approx \frac{1}{M}\sum_{m=1}^M \mathcal{S}(q(\cdot \mid y^{(m)}), \theta^{(m)}),
\end{equation}
which would guarantee $q(\theta \mid y) = \pi(\theta \mid y)$ under perfect convergence for large simulation budgets $M \rightarrow \infty$ \citep{gneiting_probabilistic_2007} and a universal density approximator $q$ \citep[e.g., coupling-based normalizing flows,][]{draxler2024universality}.
For instance, using the log-score for $S$, we retrieve the popular maximum likelihood objective,
\begin{equation}\label{eq:mle_loss}
    \mathcal{L}^{\text{MLE}}(q) := \mathbb{E}_{\pi(\theta, y)}\big[-\log q(\theta \mid y)\big].
\end{equation}
Evaluating \autoref{eq:mle_loss} empirically requires generative networks $q(\theta \mid y)$ with tractable densities, such as normalizing flows. More recently, score-based diffusion models have entered ABI \citep{sharrock2022diffusionsbi, geffner2023compositional, gloeckler2024all}, aimed at estimating the \textit{score} of the (log) posterior density $\nabla_{\theta} \log \pi(\theta \mid y)$ by minimizing a tractable re-formulation of the weighted Fisher divergence:
\begin{equation}\label{eq:score_based_fisher}
    \mathcal{L}^{\text{SM}}(s) := \frac{1}{2} \int_{0}^{T} w(t) \mathbb{E}_{\pi_t(\theta_t, y)}\left[\Vert s(\theta^{(t)}; y, t) - \nabla_{\theta} \log \pi_t(\theta^{(t)} \mid y)\Vert^2 \right]\,\diff t,
\end{equation}
where $w(t)$ is a positive weight function, $s$ is a neural network, and $\pi_t(\theta^{(t)}, y)$ is the model distribution defined in diffusion time $t$.
The most popular tractable version of \autoref{eq:score_based_fisher} is conditional denoising \citep{song2020denoising}, but the framework includes various other ``free-form'' model families capable of learning from simulations, such as flow matching \citep{albergo2023stochastic, lipman2022flow}.

Notably, it was not initially evident that neural networks could serve as standalone global posterior approximators. Early approaches, for example, combined neural networks with ABC \citep{blum2010non} or SMC \citep{paige2016inference}. \citet[p.~3]{papamakarios2016fast} noted that generative networks could be trained over the entire prior predictive distribution $\pi(y)$, but dismissed the approach as ``grossly inefficient''.

\begin{wrapfigure}[13]{R}{0.35\textwidth}
\vspace*{-1.0\baselineskip}
  \begin{center}
    \includegraphics[trim={0 14.15cm 27.05cm 0},clip,width=1.0\linewidth]{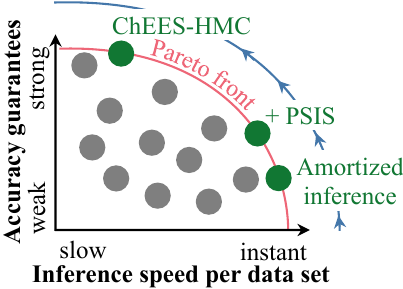}
  \end{center}
  \vspace*{-1.5\baselineskip}
  \caption{Estimation methods along the Pareto front balance a trade-off between inference speed and accuracy guarantees \citep{li2024amortized}.}
  \label{fig:pareto-front}
\end{wrapfigure}
However, subsequent work \citep{gonccalves2020training, radev2020bayesflow} showed that amortization over the prior \citep[and even across different priors,][]{elsemuller_sensitivity-aware_2024}, combined with learnable data embeddings on the fly, is in fact a viable route to fully Bayesian inference.
These developments were fueled by increasingly powerful neural density estimators \citep[e.g., coupling-based normalizing flows,][]{dinh2017density, ardizzone2018analyzing} and greater representational capacity through algorithmic alignment \citep{bloem2020probabilistic, xu2020can}.
Thus, it is now clear that amortization is not wasteful but rather essential for efficiently generating model-based insights across fields as varied as cognitive science \citep{von2022mental}, evolutionary dynamics \citep{avecilla2022neural}, and astrophysics \citep{dax2025real}, among others.

Importantly, the probabilistic calibration of all amortized methods can be efficiently checked via SBC (see \autoref{sec:model-verification:calibration} and \autoref{sec:model-checking:posterior-sbc}), since the cost of independent posterior sampling over multiple model simulations is amortized.
Additionally, the closeness of observed data to the typical set of simulations can be assessed via out-of-distribution (OOD) detection in (embedded) data space \citep{schmitt2023detecting, huang2023learning}. This is particularly important since amortized posteriors estimated from OOD data (e.g., resulting from model misspecification) tend to be biased \citep{schmitt2023detecting}, which constitutes one of the major limitations of these methods.

Amortized methods highlight a fruitful merger between scientific simulation and deep learning. Despite their advantages in big data and likelihood-free settings, they lack the guarantees of gold-standard MCMC as part of the standard Bayesian workflow \citep{gelman2020bayesian} and their accuracy can break down when asked to extrapolate beyond simulated data \citep{frazier2024statistical, schmitt2023detecting}. Recent work seeks to address these limitations through amortized workflows that combine ABI with MCMC \citep[][see \autoref{fig:pareto-front}]{li2024amortized} and trustworthy approaches that train on both simulated and real data \citep{elsemuller_does_2025, mishra_robust_2025, swierc2024domain, huang2023learning}.

\section{Model Checking}\label{sec:model-checking}
Model checking examines the compatibility of a statistical model with observed data after inference.
This \emph{data-conditional} perspective contrasts with model verification (\autoref{sec:model-verification}), which typically relies on simulated data (e.g., from the prior predictive distribution). In contrast, model checking uses the observed data $\yobs$ to assess inference validity for the case of interest. Simulations are particularly important for scaling model checking algorithms to complex, high-dimensional settings.

\subsection{Predictive Checking}
If the data generating process is formalized as a simulation program, we can generate synthetic \emph{data replications} based on the inferred parameter values.
Predictive checking \citep{Gelman2006} combines parameter estimation (inverse problem) with simulation (forward problem), thereby casting the inference results back into the data space.

\paragraph{Frequentist predictive checking} 
In the frequentist framework, predictive checking is implemented by inserting the estimated parameter values $\hat{\theta}$ into the data model $\pi(y\mid\theta)$, leading to data simulations 
\begin{equation}
    y'\sim \pi(y \mid \hat{\theta}).
\end{equation}
This process is not limited to estimates of central tendency, but it can be mirrored for other quantities, such as quantiles or confidence interval bounds of practical interest \citep{lawless2005freqpred}.
Likewise, modelers have great flexibility in the choice of metrics to evaluate the fit between the observed data $\yobs$ and the simulated data replications $y'$:
Any discrepancy measure $T(\yobs, y')$ can be used to judge whether the data replications are satisfactory (e.g., comparing averages, variances, or quantiles).

\begin{figure}[t]
    \centering
    \begin{subfigure}[b]{0.35\linewidth}
        \centering
        \includegraphics[trim={3mm 0mm 3mm 0mm},clip, width=\linewidth]{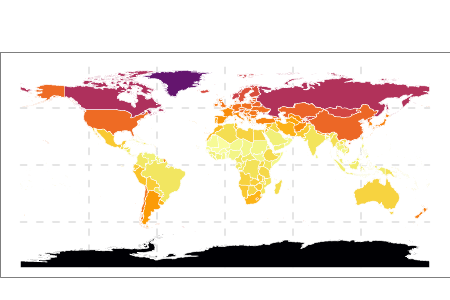}
        \includegraphics[trim={10mm 10mm 10mm 10mm},clip, width=\linewidth]{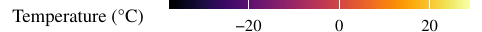}\\
        \vspace*{8mm}
        \caption{Observed data $\yobs$.}\label{fig:postpred:observed}
    \end{subfigure}
    \hfill
    \begin{subfigure}[b]{0.23\linewidth}
        \centering
        \includegraphics[trim={5mm 6mm 5mm 2mm},clip, width=\linewidth]{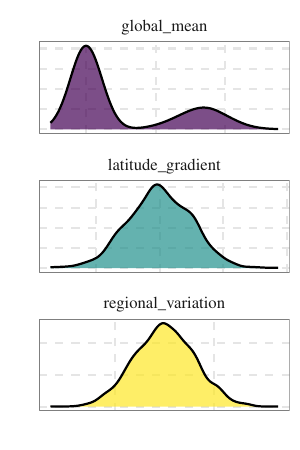}
        \caption{Posterior draws.}\label{fig:postpred:posterior}
    \end{subfigure}
    \hfill
    \begin{subfigure}[b]{0.39\linewidth}
        \centering
        \includegraphics[width=\linewidth]{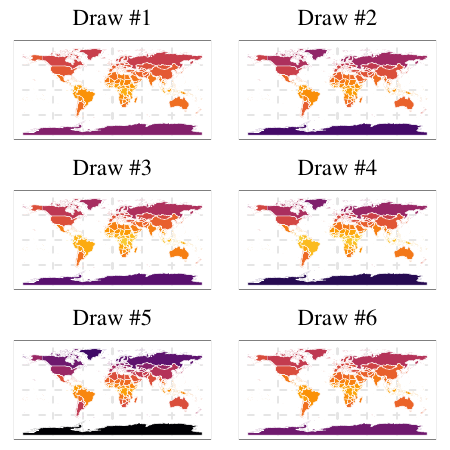}
        \caption{Posterior predictive draws.}\label{fig:postpred:posteriorpredictive}
    \end{subfigure}
    \caption{Statistical modeling of global temperature data with three parameters. 
    Based on the observed global temperature $\yobs$ (\subref{fig:postpred:observed}), Bayesian inference yields draws from the posterior distribution of the parameters $\theta$ (\subref{fig:postpred:posterior}).
    These posterior draws can be re-inserted into the simulation program (i.e., the global temperature model $f:\theta\mapsto y$) to obtain draws from the posterior predictive distribution (\subref{fig:postpred:posteriorpredictive}).
    These synthetic \emph{data replications} contain uncertainty propagated from the posterior.
    } 
    \label{fig:postpred}
\end{figure}

\paragraph{Bayesian (posterior) predictive checking} 
In the Bayesian framework, we can propagate all uncertainty from the posterior through the data model.
The result is the \emph{posterior predictive distribution}, which expresses the distribution of new data~$y'$ while accounting for all uncertainty in the system,
	\begin{equation}\label{eq:posterior-predictive}
	    \pi(y'\mid \yobs) = \int \pi(y'\mid\theta, \yobs)\, \pi(\theta\mid \yobs)\,\diff\theta.
	\end{equation}
Concretely, we can easily draw $S$ samples (i.e., data replications) from the posterior predictive distribution with an ancestral sampling scheme,
\begin{equation}\label{eq:posterior-predictive:sampling}
        \theta^{(s)}\sim \pi(\theta\mid\yobs),\quad
        y'^{(s)}_1,\ldots,y'^{(s)}_N \sim \pi(y\mid\theta^{(s)}, \yobs)\quad
        \text{for}\;s=1,\ldots,S,
\end{equation}
where each simulated data set $y'^{(s)}=\{y'^{(s)}_1,\ldots,y'^{(s)}_N\}$ contains $N$ observations.
The draws from the posterior predictive distribution represent uncertainty-aware replications of the observed data (see \autoref{fig:postpred} for an illustrative example), 
where all associated uncertainty is propagated through the forward simulation process.
Akin to frequentist predictive checking, any discrepancy statistic can then be computed from the observed data and the distribution of replicated data. 
For example, one might check whether the observed maximum value or a certain autocorrelation is extreme relative to the distribution of those metrics across the simulated replicates.
If the observed data exhibit features that would be very unlikely under the posterior predictive distribution, it indicates model misfit.

\subsection{Posterior SBC}
\label{sec:model-checking:posterior-sbc}
As detailed in \autoref{sec:model-verification:sbc}, simulation-based calibration (SBC) checks if the inference is well-calibrated for data simulated from the prior predictive distribution.
However, after observing real data $\yobs$, it is often more relevant to assess the inference conditional on that specific data rather than the entire prior predictive space.
To address this, Posterior SBC \citep{sailynoja2025posteriorsbc} uses the principles of SBC to validate the model's implementation and inference algorithm \emph{conditioned on the observed data} by means of simulations.
This approach first draws $S$ samples $\theta'^{(s)}\sim \pi(\theta\mid\yobs)$ from the (approximate) posterior conditional on the observed data $\yobs$. 
For each posterior draw, we then simulate data replications $y'^{(s)}$ from the posterior predictive distribution according to \autoref{eq:posterior-predictive:sampling}.
Finally, for each simulated data replication $y'^{(s)}$, we draw $D$ \emph{augmented} posterior samples $\theta''^{(s, 1)}, \ldots, \theta''^{(s, D)}$ conditional on the observed data $\yobs$ and the respective data replication $y'^{(s)}$,
\begin{equation}
    \theta''^{(s, 1)}, \ldots, \theta''^{(s, D)} \sim \pi(\theta\mid\yobs, y'^{(s)}),
\end{equation}
and finally check whether the augmented posterior $\pi(\theta\mid\yobs,y')$ is well-calibrated with the ``old'' posterior $\pi(\theta\mid\yobs)$ serving as prior.
Pursuing a similar data-driven perspective, \citet{fazio2024implicitpriors} propose adding small portions of real or simulated data to improve the robustness of Bayesian simulations, which in turn leads to increased stability for SBC and model simulations more broadly.

\subsection{Model comparison}
Model comparison is an umbrella term for methods that seek to evaluate multiple competing models $M_1, \ldots, M_L\in\mathcal{M}$ in the context of observed data. 
Model comparison can be framed as an extension of model checking, where instead of validating a single model in isolation with respect to some metric, we assess \emph{multiple candidate models simultaneously} using suitable metrics that enable direct comparisons.

Formalizing the process of comparing models helps researchers make principled and reproducible decisions about model selection, and this process can be supported with model simulations.
In the context of model comparison, simulations serve a dual purpose:
First, simulations render otherwise intractable algorithms computationally feasible (e.g., marginal likelihood estimation with neural simulation-based inference).
Second, simulations constitute the core principle of model comparison methods, such as predictive checking of multiple candidate models or meta-uncertainty (see below).

Within the scope of model simulations for Bayesian inference, model comparison extends the data-generating simulation process by explicitly encoding the statistical data-generating model $M_l$,
\begin{equation}
    \yobs\sim \pi(y\mid\theta, M_l) \;\text{with}\; \theta\sim \pi(\theta\mid M_l) \;\text{and}\; M_l\sim \pi(M),
\end{equation}
where $\pi(M)$ is a discrete prior distribution over the candidate models.
In \emph{prior-based model comparison}, the central quantity is the marginal likelihood $\pi(\yobs\mid M_l) = \int \pi(\yobs\mid\theta, M_l)\,\pi(\theta\mid M_l)\diff\theta$, which quantifies the evidence for model $M_l$ by integrating over all parameter values.
This enables the computation of Bayes Factors \citep{kass_bayes_1995} and posterior model probabilities.
Since computing the marginal likelihood requires solving a potentially high-dimensional integral, this family of model comparison methods is typically computationally infeasible for complex models.
Based on synthetic simulations from the joint model $\pi(\theta, y)$, amortized inference tackles this problem by directly learning the evidence \citep[aka.\ evidential learning;][]{radev_amortized_2021} or by simultaneously learning an amortized posterior and likelihood approximator \citep{radev_jana_2023}.

By design, prior-based model comparison is sensitive to the choice of prior \citep{Schad2023, oelrich_when_2020}, which often leads to issues in practical scenarios where prior specification is a hard challenge \citep{Aguilar2023,dellaportas_joint_2012,van_dongen_prior_2006,zitzmann_prior_2021,lindley_statistical_1957}.
Differently, \emph{posterior-based methods}, such as cross-validation (CV), estimate the \textit{expected} predictive distribution over new data, $
\mathbb{E}\left[\pi(y' \mid \yobs, M_l)\right]$ \citep{vehtari_practical_2017, vehtari2019limitations}. 
Taken to the extreme, leave-one-out cross-validation (LOO-CV) evaluates the predictive distribution of single held-out observations \citep{vehtari_practical_2017}.
Importantly, amortized methods can perform overhead-free LOO-CV (or other cross-validation schemes) in cases where established approximate methods \citep[e.g., Pareto-smoothed importance sampling,][]{vehtari2024pareto} are infeasible \citep{radev_jana_2023}.

While less sensitive to prior specification, posterior-based model comparison methods lack the consistency guarantees of marginal likelihood approaches, that is, they are not bound to recover the true data-generating model in the asymptotic limit of infinite data.
Here again, model simulations have the potential to improve what we can learn from a limited amount of observed data:
For example, \cite{schmitt_meta-uncertainty_2023} developed a meta-uncertainty framework to judge the \emph{replicability} of Bayesian model comparison by combining (i) prior-based model comparison; (ii) frequentist sampling distributions based on model simulations; and (iii) posterior predictive distributions based on observed data.

\subsection{Sensitivity analysis}\label{sec:model-checking:sensitivity}

Sensitivity analysis \citep[aka.\ robustness analysis;][]{Berger1994} constitutes a critical step in statistical modeling that focuses on understanding how the output of a model changes in response to variations in its inputs. 
These inputs can include fixed (hyper-)parameters of the model, the specification of prior distributions and observation models in a Bayesian setting, or the choices made while pre-processing the data. 
As such, each analysis \emph{hides an iceberg of uncertainty} \citep{Wagenmakers2022}, and sensitivity analysis is a principled approach to render this uncertainty tangible.
In the PAD taxonomy (see \autoref{sec:model-specification}), sensitivity analyses can target each of the three axes of Bayesian models: Different statistical models, varying posterior approximators, and different variations of the training data. 

Since the assumed data-generating process is essentially a simulation program, simulations also lie at the heart of sensitivity analysis.
By systematically varying the inputs of an analysis within a plausible space, then executing the simulation program, and ultimately observing the resulting changes in the outputs, researchers can assess the robustness of their conclusions to these initial variations \citep{Berger1994}. 
Simulation represents the bridging element between input variations on the one hand and inference results on the other hand.
At the same time, hyperparameters in the simulation process itself can be subject to sensitivity analyses as well, which allows for powerful conclusions about the data-generating process but also comes with a yet increased degree of complexity.
While sensitivity analyses might seem fairly straightforward for trivial models with low complexity, the computational demands for principled sensitivity analyses quickly become unbearable for complex statistical analyses with high-dimensional parameter spaces and involved data pre-processing routines.

Akin to tracing all sources of uncertainty in Bayesian analysis, keeping track of all possible choices in sensitivity analysis resembles a \emph{Garden of Forking Paths} \citep{borges_garden_1941,mcelreath_statistical_2020}.
Taming this combinatorial explosion is the key to enabling large-scale sensitivity-analyses with finite computational resources.
To this end, \citet{Kallioinen2023} developed a computationally efficient approach that uses importance sampling to estimate the effect of power-scaling the prior and likelihood in a Bayesian analysis.
\citet{elsemuller_sensitivity-aware_2024} proposed \emph{sensitivity-aware amortized Bayesian inference}, which extends the scope of amortization by encoding context information about the statistical model.
By additionally conditioning on this context information during the simulation-based training stage, the trained neural networks can subsequently perform near-instant sensitivity analyses without worrying about the associated cost of modifying model assumptions during inference.

\subsection{Increasing the efficiency of simulation-based estimators}

Many use cases of simulations we highlighted above apply the standard Monte Carlo (MC) estimator $S^{-1} \sum_{s=1} f(\theta^{(s)})$ for approximating a target expectation $\mathbb{E}[f(\theta)]$. However, simulation-based estimators can become more sample-efficient than the standard MC estimator if additional information is available and correctly utilized. For example, in latent variable models, analytically integrating out the latent variables (if possible) before performing predictive simulations on new data can increase the stability of the resulting estimators \citep{merkle2019bayesian}. As another example, ``control variates'' can improve the efficiency of Monte Carlo estimators by leveraging the score (i.e., the gradient of the log target density) when available \citep{south2023regularized}.
More generally, if conditional expectations of the form $\mathbb{E}[f(\theta) \mid T(y)]$ for an informative statistic $T(y)$ are available, approximating these conditional expectations via simulations can yield a more efficient estimator of $\mathbb{E}[f(\theta)]$, a process also known as Rao-Blackwellization \citep[see][for an overview]{robert2021rao}. These examples illustrate that simulation-based methods are not an orthogonal alternative to analytical or density-based methods but can also benefit from the latter provided the right type of additional information.

\section{Conclusion and Outlook}

In the preceding pages, we have seen that simulations are a powerful and versatile tool throughout all major steps of statistical workflows. In the foreseeable future, we expect the utility and use of simulations to increase even further. In particular, amortized inference, which relies heavily on both model simulations and continuous progress in deep learning, is likely to become a viable, widespread alternative to established inference approaches. 

Going beyond the techniques available to date, one could imagine simulation-based training of entire world models that are able to rapidly and semi-automatically execute full statistical workflows on real data, essentially playing the role of a statistics expert assisting the user in their analysis. One of the key challenges for such machine-assisted statistics is how to combine the general expertise of the machine with the subject matter knowledge of the user. No matter how many simulations the machine has been trained on, it would still likely miss key subject matter knowledge about the specific real data being analyzed, just as a statistics expert wouldn't know all the intricate details of the data. 

This also highlights a more general question pertinent to all inference machines trained on simulated data: how to bridge the statistical gap between simulated worlds and the real world. Put differently, we need to learn how to formally integrate information from simulated data, real data, and subject matter knowledge into inferences that are both fast and trustworthy. Much remains to be simulated.

\newpage

\section*{Acknowledgements}

Paul Bürkner acknowledges support of the Deutsche Forschungsgemeinschaft (DFG, German Research Foundation) via Projects 508399956 and 528702768 as well as the Collaborative Research Center 391 (Spatio-Temporal Statistics for the Transition of Energy and Transport) – 520388526. Stefan Radev is supported by the National Science Foundation under Grant No.~2448380.

\bibliographystyle{apalike}
\bibliography{references}

\begin{thebibliography}{}

\bibitem[Aguilar and B\"{u}rkner, 2023]{Aguilar2023}
Aguilar, J.~E. and B\"{u}rkner, P.-C. (2023).
\newblock Intuitive joint priors for {Bayesian} linear multilevel models: The {R2D2M2} prior.
\newblock {\em Electronic Journal of Statistics}, 17(1).

\bibitem[Aguilera-Venegas et~al., 2019]{aguilera2019probabilistic}
Aguilera-Venegas, G., Gal{\'a}n-Garc{\'\i}a, J.~L., Egea-Guerrero, R., Gal{\'a}n-Garc{\'\i}a, M.~{\'A}., Rodr{\'\i}guez-Cielos, P., Padilla-Dom{\'\i}nguez, Y., and Gal{\'a}n-Luque, M. (2019).
\newblock A probabilistic extension to conway’s game of life.
\newblock {\em Advances in Computational Mathematics}, 45:2111--2121.

\bibitem[Albergo et~al., 2023]{albergo2023stochastic}
Albergo, M.~S., Boffi, N.~M., and Vanden-Eijnden, E. (2023).
\newblock Stochastic interpolants: A unifying framework for flows and diffusions.
\newblock {\em arXiv preprint:2303.08797}.

\bibitem[Ardizzone et~al., 2018]{ardizzone2018analyzing}
Ardizzone, L., Kruse, J., Rother, C., and K{\"o}the, U. (2018).
\newblock Analyzing inverse problems with invertible neural networks.
\newblock In {\em International Conference on Learning Representations}.

\bibitem[Avecilla et~al., 2022]{avecilla2022neural}
Avecilla, G., Chuong, J.~N., Li, F., Sherlock, G., Gresham, D., and Ram, Y. (2022).
\newblock Neural networks enable efficient and accurate simulation-based inference of evolutionary parameters from adaptation dynamics.
\newblock {\em PLoS biology}, 20(5):e3001633.

\bibitem[Bansal et~al., 2025]{bansal2025surprising}
Bansal, V., Chen, T., and Scott, J.~G. (2025).
\newblock The surprising strength of weak classifiers for validating neural posterior estimates.
\newblock {\em arXiv preprint arXiv:2507.17026}.

\bibitem[Beaumont et~al., 2009]{beaumont2009adaptive}
Beaumont, M.~A., Cornuet, J.-M., Marin, J.-M., and Robert, C.~P. (2009).
\newblock Adaptive approximate {B}ayesian computation.
\newblock {\em Biometrika}, 96(4):983--990.

\bibitem[Berger et~al., 1994]{Berger1994}
Berger, J.~O., Moreno, E., Pericchi, L.~R., Bayarri, M.~J., Bernardo, J.~M., Cano, J.~A., De~la Horra, J., Martín, J., Ríos-Insúa, D., Betrò, B., Dasgupta, A., Gustafson, P., Wasserman, L., Kadane, J.~B., Srinivasan, C., Lavine, M., O’Hagan, A., Polasek, W., Robert, C.~P., Goutis, C., Ruggeri, F., Salinetti, G., and Sivaganesan, S. (1994).
\newblock An overview of robust {Bayesian} analysis.
\newblock {\em Test}, 3(1):5–124.

\bibitem[Biron-Lattes et~al., 2024]{biron2024automala}
Biron-Lattes, M., Surjanovic, N., Syed, S., Campbell, T., and Bouchard-C{\^o}t{\'e}, A. (2024).
\newblock automala: Locally adaptive metropolis-adjusted langevin algorithm.
\newblock In {\em International Conference on Artificial Intelligence and Statistics}, pages 4600--4608. PMLR.

\bibitem[Bloem-Reddy and Teh, 2020]{bloem2020probabilistic}
Bloem-Reddy, B. and Teh, Y.~W. (2020).
\newblock Probabilistic symmetries and invariant neural networks.
\newblock {\em Journal of Machine Learning Research}, 21(90):1--61.

\bibitem[Blum and Fran{\c{c}}ois, 2010]{blum2010non}
Blum, M.~G. and Fran{\c{c}}ois, O. (2010).
\newblock Non-linear regression models for approximate {Bayesian} computation.
\newblock {\em Statistics and computing}, 20:63--73.

\bibitem[Bockting et~al., 2024a]{bockting2024expert}
Bockting, F., Radev, S.~T., and B{\"u}rkner, P.-C. (2024a).
\newblock Expert-elicitation method for non-parametric joint priors using normalizing flows.
\newblock {\em arXiv preprint arXiv:2411.15826}.

\bibitem[Bockting et~al., 2024b]{bockting2024simulation}
Bockting, F., Radev, S.~T., and B{\"u}rkner, P.-C. (2024b).
\newblock Simulation-based prior knowledge elicitation for parametric {B}ayesian models.
\newblock {\em Scientific Reports}, 14(1):17330.

\bibitem[Borges, 1941]{borges_garden_1941}
Borges, J.~L. (1941).
\newblock {\em Garden of {Forking} {Paths}}.
\newblock n.p.

\bibitem[Brooks et~al., 2011]{brooks2011handbook}
Brooks, S., Gelman, A., Jones, G., and Meng, X.-L. (2011).
\newblock {\em Handbook of Markov chain Monte Carlo}.
\newblock CRC press.

\bibitem[Bürkner et~al., 2023]{burkner_models_2023}
Bürkner, P.-C., Scholz, M., and Radev, S.~T. (2023).
\newblock Some models are useful, but how do we know which ones? {Towards} a unified {Bayesian} model taxonomy.
\newblock {\em Statistics Surveys}, 17:216--310.

\bibitem[Casti, 1996]{casti1996would}
Casti, J.~L. (1996).
\newblock {\em Would-be worlds: How simulation is changing the frontiers of science}.
\newblock John Wiley \& Sons, Inc.

\bibitem[Cohen, 2013]{cohen_statistical_2013}
Cohen, J. (2013).
\newblock {\em Statistical {Power} {Analysis} for the {Behavioral} {Sciences}}.
\newblock Routledge, 2 edition.

\bibitem[Cook et~al., 2006]{cook2006validation}
Cook, S.~R., Gelman, A., and Rubin, D.~B. (2006).
\newblock Validation of software for bayesian models using posterior quantiles.
\newblock {\em Journal of Computational and Graphical Statistics}, 15(3):675--692.

\bibitem[Cranmer et~al., 2020]{cranmer2020frontier}
Cranmer, K., Brehmer, J., and Louppe, G. (2020).
\newblock The frontier of simulation-based inference.
\newblock {\em Proceedings of the National Academy of Sciences}, 117(48):30055--30062.

\bibitem[Dalmasso et~al., 2024]{dalmasso2024likelihood}
Dalmasso, N., Masserano, L., Zhao, D., Izbicki, R., and Lee, A.~B. (2024).
\newblock Likelihood-free frequentist inference: Bridging classical statistics and machine learning for reliable simulator-based inference.
\newblock {\em Electronic Journal of Statistics}, 18(2):5045--5090.

\bibitem[Dax et~al., 2025]{dax2025real}
Dax, M., Green, S.~R., Gair, J., Gupte, N., P{\"u}rrer, M., Raymond, V., Wildberger, J., Macke, J.~H., Buonanno, A., and Sch{\"o}lkopf, B. (2025).
\newblock Real-time inference for binary neutron star mergers using machine learning.
\newblock {\em Nature}, 639(8053):49--53.

\bibitem[Del~Moral et~al., 2006]{del2006sequential}
Del~Moral, P., Doucet, A., and Jasra, A. (2006).
\newblock Sequential {Monte} {Carlo} samplers.
\newblock {\em Journal of the Royal Statistical Society Series B: Statistical Methodology}, 68(3):411--436.

\bibitem[Del~Moral et~al., 2012]{del2012adaptive}
Del~Moral, P., Doucet, A., and Jasra, A. (2012).
\newblock An adaptive sequential {Monte Carlo method for approximate Bayesian computation}.
\newblock {\em Statistics and Computing}, 22(5):1009--1020.

\bibitem[Dellaportas et~al., 2012]{dellaportas_joint_2012}
Dellaportas, P., Forster, J.~J., and Ntzoufras, I. (2012).
\newblock Joint {Specification} of {Model} {Space} and {Parameter} {Space} {Prior} {Distributions}.
\newblock {\em Statistical Science}, 27(2).

\bibitem[Diggle and Gratton, 1984]{diggle1984monte}
Diggle, P.~J. and Gratton, R.~J. (1984).
\newblock Monte carlo methods of inference for implicit statistical models.
\newblock {\em Journal of the Royal Statistical Society Series B: Statistical Methodology}, 46(2):193--212.

\bibitem[Dinh et~al., 2017]{dinh2017density}
Dinh, L., Sohl-Dickstein, J., and Bengio, S. (2017).
\newblock Density estimation using real nvp.
\newblock In {\em International Conference on Learning Representations}.

\bibitem[Draxler et~al., 2024]{draxler2024universality}
Draxler, F., Wahl, S., Schn{\"o}rr, C., and K{\"o}the, U. (2024).
\newblock On the universality of coupling-based normalizing flows.
\newblock {\em arXiv preprint}.

\bibitem[Dur{\'a}n, 2020]{duran2020simulation}
Dur{\'a}n, J.~M. (2020).
\newblock What is a simulation model?
\newblock {\em Minds and Machines}, 30(3):301--323.

\bibitem[Efron and Hastie, 2021]{efron2021computer}
Efron, B. and Hastie, T. (2021).
\newblock {\em Computer age statistical inference, student edition: algorithms, evidence, and data science}, volume~6.
\newblock Cambridge University Press.

\bibitem[Elsemüller et~al., 2024]{elsemuller_sensitivity-aware_2024}
Elsemüller, L., Olischläger, H., Schmitt, M., Bürkner, P.-C., Koethe, U., and Radev, S.~T. (2024).
\newblock Sensitivity-aware amortized {Bayesian} inference.
\newblock {\em Transactions on Machine Learning Research}.

\bibitem[Elsemüller et~al., 2025]{elsemuller_does_2025}
Elsemüller, L., Pratz, V., Krause, M.~v., Voss, A., Bürkner, P.-C., and Radev, S.~T. (2025).
\newblock Does {Unsupervised} {Domain} {Adaptation} {Improve} the {Robustness} of {Amortized} {Bayesian} {Inference}? {A} {Systematic} {Evaluation}.
\newblock {\em arXiv preprint}.

\bibitem[Fazio et~al., 2024]{fazio2024implicitpriors}
Fazio, L., Scholz, M., and Bürkner, P.-C. (2024).
\newblock Primed priors for simulation-based validation of {Bayesian} models.
\newblock {\em arXiv preprint:2408.06504}.

\bibitem[Frazier et~al., 2024]{frazier2024statistical}
Frazier, D.~T., Kelly, R., Drovandi, C., and Warne, D.~J. (2024).
\newblock The statistical accuracy of neural posterior and likelihood estimation.
\newblock {\em arXiv preprint arXiv:2411.12068}.

\bibitem[Frazier et~al., 2018]{frazier2018asymptotic}
Frazier, D.~T., Martin, G.~M., Robert, C.~P., and Rousseau, J. (2018).
\newblock Asymptotic properties of approximate {Bayesian} computation.
\newblock {\em Biometrika}, 105(3):593--607.

\bibitem[Gabry et~al., 2019]{gabry2019visualization}
Gabry, J., Simpson, D., Vehtari, A., Betancourt, M., and Gelman, A. (2019).
\newblock Visualization in bayesian workflow.
\newblock {\em Journal of the Royal Statistical Society Series A: Statistics in Society}, 182(2):389--402.

\bibitem[Geffner et~al., 2023]{geffner2023compositional}
Geffner, T., Papamakarios, G., and Mnih, A. (2023).
\newblock Compositional score modeling for simulation-based inference.
\newblock In {\em International Conference on Machine Learning}, pages 11098--11116. PMLR.

\bibitem[Gelfand and Smith, 1990]{gelfand1990sampling}
Gelfand, A.~E. and Smith, A.~F. (1990).
\newblock Sampling-based approaches to calculating marginal densities.
\newblock {\em Journal of the American statistical association}, 85(410):398--409.

\bibitem[Gelman et~al., 2013]{gelman_bayesian_2013}
Gelman, A., Carlin, J.~B., Stern, H.~S., Dunson, D.~B., Vehtari, A., and Rubin, D.~B. (2013).
\newblock {\em Bayesian {Data} {Analysis} (3rd {Edition})}.
\newblock London: Chapman and Hall/CRC.

\bibitem[Gelman and Hill, 2006]{Gelman2006}
Gelman, A. and Hill, J. (2006).
\newblock {\em Analytical methods for social research: Data analysis using regression and multilevel/hierarchical models}.
\newblock Cambridge University Press, Cambridge, England.

\bibitem[Gelman et~al., 2020]{gelman2020bayesian}
Gelman, A., Vehtari, A., Simpson, D., Margossian, C.~C., Carpenter, B., Yao, Y., Kennedy, L., Gabry, J., B{\"u}rkner, P.-C., and Modr{\'a}k, M. (2020).
\newblock Bayesian workflow.
\newblock {\em arXiv preprint:2011.01808}.

\bibitem[Geman and Geman, 1984]{geman1984stochastic}
Geman, S. and Geman, D. (1984).
\newblock Stochastic relaxation, gibbs distributions, and the bayesian restoration of images.
\newblock {\em IEEE Transactions on Pattern Analysis and Machine Intelligence}, 6:721--741.

\bibitem[Gershman and Goodman, 2014]{gershman2014amortized}
Gershman, S. and Goodman, N. (2014).
\newblock Amortized inference in probabilistic reasoning.
\newblock In {\em Proceedings of the annual meeting of the cognitive science society}, volume~36.

\bibitem[Gloeckler et~al., 2024]{gloeckler2024all}
Gloeckler, M., Deistler, M., Weilbach, C., Wood, F., and Macke, J.~H. (2024).
\newblock All-in-one simulation-based inference.
\newblock {\em arXiv preprint}.

\bibitem[Gneiting et~al., 2007]{gneiting_probabilistic_2007}
Gneiting, T., Balabdaoui, F., and Raftery, A.~E. (2007).
\newblock Probabilistic {Forecasts}, {Calibration} and {Sharpness}.
\newblock {\em Journal of the Royal Statistical Society Series B: Statistical Methodology}, 69(2):243--268.

\bibitem[Gon{\c{c}}alves et~al., 2020]{gonccalves2020training}
Gon{\c{c}}alves, P.~J., Lueckmann, J.-M., Deistler, M., Nonnenmacher, M., {\"O}cal, K., Bassetto, G., Chintaluri, C., Podlaski, W.~F., Haddad, S.~A., Vogels, T.~P., et~al. (2020).
\newblock Training deep neural density estimators to identify mechanistic models of neural dynamics.
\newblock {\em elife}, 9:e56261.

\bibitem[Good, 1950]{good1950probability}
Good, I.~J. (1950).
\newblock {\em Probability and the Weighing of Evidence}.
\newblock New York: Hafners.

\bibitem[Grim and Singer, 2024]{grim2024comp}
Grim, P. and Singer, D. (2024).
\newblock Computational philosophy.
\newblock In Zalta, E.~N. and Nodelman, U., editors, {\em The Stanford Encyclopedia of Philosophy (Summer 2024 Edition)}. Metaphysics Research Lab, Stanford University.

\bibitem[Guala, 2002]{guala2002models}
Guala, F. (2002).
\newblock Models, simulations, and experiments.
\newblock In {\em Model-based reasoning: Science, technology, values}, pages 59--74. Springer.

\bibitem[Hartmann et~al., 2020]{hartmann2020flexible}
Hartmann, M., Agiashvili, G., B{\"u}rkner, P., and Klami, A. (2020).
\newblock Flexible prior elicitation via the prior predictive distribution.
\newblock In {\em Conference on Uncertainty in Artificial Intelligence}, pages 1129--1138. PMLR.

\bibitem[Hastings, 1970]{hastings1970monte}
Hastings, W.~K. (1970).
\newblock Monte carlo sampling methods using markov chains and their applications.
\newblock {\em Biometrika}, 57(1):97--109.

\bibitem[Hermans et~al., 2020]{hermans2020likelihood}
Hermans, J., Begy, V., and Louppe, G. (2020).
\newblock Likelihood-free mcmc with amortized approximate ratio estimators.
\newblock In {\em International conference on machine learning}, pages 4239--4248. PMLR.

\bibitem[Huang et~al., 2023]{huang2023learning}
Huang, D., Bharti, A., Souza, A., Acerbi, L., and Kaski, S. (2023).
\newblock Learning robust statistics for simulation-based inference under model misspecification.
\newblock {\em Advances in Neural Information Processing Systems}, 36:7289--7310.

\bibitem[Ibragimov and Has'minskii, 2013]{ibragimov_statistical_2013}
Ibragimov, I.~A. and Has'minskii, R.~Z. (2013).
\newblock {\em Statistical {Estimation}: {Asymptotic} {Theory}}.
\newblock Springer Science \& Business Media.

\bibitem[Kallioinen et~al., 2023]{Kallioinen2023}
Kallioinen, N., Paananen, T., B\"{u}rkner, P.-C., and Vehtari, A. (2023).
\newblock Detecting and diagnosing prior and likelihood sensitivity with power-scaling.
\newblock {\em Statistics and Computing}, 34(1).

\bibitem[Kass and Raftery, 1995]{kass_bayes_1995}
Kass, R.~E. and Raftery, A.~E. (1995).
\newblock Bayes {Factors}.
\newblock {\em Journal of the American Statistical Association}, 90(430):773--795.

\bibitem[Kingma and Welling, 2013]{kingma2013auto}
Kingma, D.~P. and Welling, M. (2013).
\newblock Auto-encoding variational bayes.
\newblock {\em arXiv preprint arXiv:1312.6114}.

\bibitem[Kucharski, 2016]{kucharski2016perfect}
Kucharski, A. (2016).
\newblock {\em The perfect bet: How science and math are taking the luck out of gambling}.
\newblock Hachette UK.

\bibitem[Lavin et~al., 2021]{lavin2021simulation}
Lavin, A., Krakauer, D., Zenil, H., Gottschlich, J., Mattson, T., Brehmer, J., Anandkumar, A., Choudry, S., Rocki, K., Baydin, A.~G., et~al. (2021).
\newblock Simulation intelligence: Towards a new generation of scientific methods.
\newblock {\em arXiv preprint}.

\bibitem[Lawless and Fredette, 2005]{lawless2005freqpred}
Lawless, J.~F. and Fredette, M. (2005).
\newblock Frequentist prediction intervals and predictive distributions.
\newblock {\em Biometrika}, 92(3):529--542.

\bibitem[Le et~al., 2017]{le2017inference}
Le, T.~A., Baydin, A.~G., and Wood, F. (2017).
\newblock Inference compilation and universal probabilistic programming.
\newblock In {\em Artificial Intelligence and Statistics}, pages 1338--1348. PMLR.

\bibitem[Li et~al., 2024]{li2024amortized}
Li, C., Vehtari, A., B{\"u}rkner, P.-C., Radev, S.~T., Acerbi, L., and Schmitt, M. (2024).
\newblock Amortized bayesian workflow.
\newblock {\em arXiv preprint arXiv:2409.04332}.

\bibitem[Lindley, 1957]{lindley_statistical_1957}
Lindley, D.~V. (1957).
\newblock A {Statistical} {Paradox}.
\newblock {\em Biometrika}, 44(1-2):187--192.

\bibitem[Lipman et~al., 2022]{lipman2022flow}
Lipman, Y., Chen, R.~T., Ben-Hamu, H., Nickel, M., and Le, M. (2022).
\newblock Flow matching for generative modeling.
\newblock {\em arXiv preprint:2210.02747}.

\bibitem[Little, 2006]{little_calibrated_2006}
Little, R.~J. (2006).
\newblock Calibrated {Bayes}.
\newblock {\em The American Statistician}, 60(3):213--223.

\bibitem[MacKay, 2003]{mackay2003information}
MacKay, D.~J. (2003).
\newblock {\em Information theory, inference and learning algorithms}.
\newblock Cambridge university press.

\bibitem[Margossian et~al., 2024]{margossian2024nested}
Margossian, C.~C., Hoffman, M.~D., Sountsov, P., Riou-Durand, L., Vehtari, A., and Gelman, A. (2024).
\newblock Nested ˆr: Assessing the convergence of markov chain monte carlo when running many short chains.
\newblock {\em Bayesian Analysis}, 1(1):1--28.

\bibitem[Marin et~al., 2012]{marin2012approximate}
Marin, J.-M., Pudlo, P., Robert, C.~P., and Ryder, R.~J. (2012).
\newblock Approximate {B}ayesian computational methods.
\newblock {\em Statistics and Computing}, 22(6):1167--1180.

\bibitem[Marjoram et~al., 2003]{marjoram2003markov}
Marjoram, P., Molitor, J., Plagnol, V., and Tavar{\'e}, S. (2003).
\newblock Markov chain {M}onte {C}arlo without likelihoods.
\newblock {\em Proceedings of the National Academy of Sciences}, 100(26):15324--15328.

\bibitem[McElreath, 2020]{mcelreath_statistical_2020}
McElreath, R. (2020).
\newblock {\em Statistical rethinking: a {Bayesian} course with examples in {R} and {Stan}}.
\newblock Chapman \& {Hall}/{CRC} texts in statistical science series. CRC Press, Taylor \& Francis Group, Boca Raton London New York, second edition edition.

\bibitem[Merkle et~al., 2019]{merkle2019bayesian}
Merkle, E.~C., Furr, D., and Rabe-Hesketh, S. (2019).
\newblock Bayesian comparison of latent variable models: Conditional versus marginal likelihoods.
\newblock {\em Psychometrika}, 84(3):802--829.

\bibitem[Metropolis et~al., 1953]{metropolis1953equation}
Metropolis, N., Rosenbluth, A.~W., Rosenbluth, M.~N., Teller, A.~H., and Teller, E. (1953).
\newblock Equation of state calculations by fast computing machines.
\newblock {\em The journal of chemical physics}, 21(6):1087--1092.

\bibitem[Mikkola et~al., 2024a]{mikkola2024preferential}
Mikkola, P., Acerbi, L., and Klami, A. (2024a).
\newblock Preferential normalizing flows.
\newblock {\em arXiv preprint}.

\bibitem[Mikkola et~al., 2024b]{mikkola2024prior}
Mikkola, P., Martin, O.~A., Chandramouli, S., Hartmann, M., Abril~Pla, O., Thomas, O., Pesonen, H., Corander, J., Vehtari, A., Kaski, S., et~al. (2024b).
\newblock Prior knowledge elicitation: The past, present, and future.
\newblock {\em Bayesian Analysis}, 19(4):1129--1161.

\bibitem[Mishra et~al., 2025]{mishra_robust_2025}
Mishra, A., Habermann, D., Schmitt, M., Radev, S.~T., and Bürkner, P.-C. (2025).
\newblock Robust {Amortized} {Bayesian} {Inference} with {Self}-{Consistency} {Losses} on {Unlabeled} {Data}.
\newblock {\em arXiv preprint}.

\bibitem[Modi et~al., 2024]{modi2024delayed}
Modi, C., Barnett, A., and Carpenter, B. (2024).
\newblock Delayed rejection hamiltonian monte carlo for sampling multiscale distributions.
\newblock {\em Bayesian Analysis}, 19(3):815--842.

\bibitem[Modr{\'a}k et~al., 2025]{modrak2025simulation}
Modr{\'a}k, M., Moon, A.~H., Kim, S., B{\"u}rkner, P., Huurre, N., Faltejskov{\'a}, K., Gelman, A., and Vehtari, A. (2025).
\newblock Simulation-based calibration checking for bayesian computation: The choice of test quantities shapes sensitivity.
\newblock {\em Bayesian Analysis}, 20(2):461--488.

\bibitem[Oelrich et~al., 2020]{oelrich_when_2020}
Oelrich, O., Ding, S., Magnusson, M., Vehtari, A., and Villani, M. (2020).
\newblock When are {Bayesian} model probabilities overconfident?
\newblock {\em arXiv preprint:2003.04026}.

\bibitem[Paige and Wood, 2016]{paige2016inference}
Paige, B. and Wood, F. (2016).
\newblock Inference networks for sequential monte carlo in graphical models.
\newblock In {\em International Conference on Machine Learning}, pages 3040--3049. PMLR.

\bibitem[Papamakarios and Murray, 2016]{papamakarios2016fast}
Papamakarios, G. and Murray, I. (2016).
\newblock Fast $\varepsilon$-free inference of simulation models with {Bayesian} conditional density estimation.
\newblock {\em Advances in neural information processing systems}, 29.

\bibitem[Peskun, 1973]{peskun1973optimum}
Peskun, P.~H. (1973).
\newblock Optimum monte-carlo sampling using markov chains.
\newblock {\em Biometrika}, 60(3):607--612.

\bibitem[Picchini, 2014]{picchini2014inference}
Picchini, U. (2014).
\newblock Inference for {SDE} models via approximate {Bayesian} computation.
\newblock {\em Journal of Computational and Graphical Statistics}, 23(4):1080--1100.

\bibitem[Picchini and Tamborrino, 2024]{picchini2024guided}
Picchini, U. and Tamborrino, M. (2024).
\newblock Guided sequential abc schemes for intractable {Bayesian} models.
\newblock {\em Bayesian Analysis}, 1(1):1--32.

\bibitem[Racine and Mackinnon, 2007]{racine_simulation-based_2007}
Racine, J.~S. and Mackinnon, J.~G. (2007).
\newblock Simulation-{Based} {Tests} that {Can} {Use} {Any} {Number} of {Simulations}.
\newblock {\em Communications in Statistics - Simulation and Computation}, 36(2):357--365.

\bibitem[Radev et~al., 2021]{radev_amortized_2021}
Radev, S.~T., D’Alessandro, M., Mertens, U.~K., Voss, A., Köthe, U., and Bürkner, P.-C. (2021).
\newblock Amortized {Bayesian} {Model} {Comparison} {With} {Evidential} {Deep} {Learning}.
\newblock {\em IEEE Transactions on Neural Networks and Learning Systems}, 34(8):4903--4917.

\bibitem[Radev et~al., 2020]{radev2020bayesflow}
Radev, S.~T., Mertens, U.~K., Voss, A., Ardizzone, L., and K{\"o}the, U. (2020).
\newblock Bayesflow: Learning complex stochastic models with invertible neural networks.
\newblock {\em IEEE Transactions on Neural Networks and Learning Systems}, 33(4):1452--1466.

\bibitem[Radev et~al., 2023]{radev_jana_2023}
Radev, S.~T., Schmitt, M., Pratz, V., Picchini, U., Köthe, U., and Bürkner, P.-C. (2023).
\newblock {JANA}: {Jointly} amortized neural approximation of complex {Bayesian} models.
\newblock In Evans, R.~J. and Shpitser, I., editors, {\em Proceedings of the thirty-ninth conference on uncertainty in artificial intelligence}, volume 216 of {\em Proceedings of machine learning research}, pages 1695--1706. PMLR.

\bibitem[Rezende et~al., 2014]{rezende2014stochastic}
Rezende, D.~J., Mohamed, S., and Wierstra, D. (2014).
\newblock Stochastic backpropagation and approximate inference in deep generative models.
\newblock In {\em International conference on machine learning}, pages 1278--1286. PMLR.

\bibitem[Robert and Casella, 2011]{robert2011short}
Robert, C. and Casella, G. (2011).
\newblock A short history of markov chain monte carlo: Subjective recollections from incomplete data.
\newblock {\em Statistical Science}, pages 102--115.

\bibitem[Robert and Roberts, 2021]{robert2021rao}
Robert, C.~P. and Roberts, G.~O. (2021).
\newblock Rao-blackwellization in the mcmc era.
\newblock {\em arXiv preprint:2101.01011}.

\bibitem[Rosenblatt, 1952]{rosenblatt_remarks_1952}
Rosenblatt, M. (1952).
\newblock Remarks on a {Multivariate} {Transformation}.
\newblock {\em The Annals of Mathematical Statistics}, 23(3):470--472.

\bibitem[Rubin, 1984]{rubin1984bayesianly}
Rubin, D.~B. (1984).
\newblock Bayesianly justifiable and relevant frequency calculations for the applied statistician.
\newblock {\em The Annals of Statistics}, pages 1151--1172.

\bibitem[Schad et~al., 2021]{schad2021toward}
Schad, D.~J., Betancourt, M., and Vasishth, S. (2021).
\newblock Toward a principled {B}ayesian workflow in cognitive science.
\newblock {\em Psychological methods}, 26(1):103.

\bibitem[Schad et~al., 2023]{Schad2023}
Schad, D.~J., Nicenboim, B., B\"{u}rkner, P.-C., Betancourt, M., and Vasishth, S. (2023).
\newblock Workflow techniques for the robust use of bayes factors.
\newblock {\em Psychological Methods}, 28(6):1404–1426.

\bibitem[Schmitt et~al., 2023a]{schmitt2023detecting}
Schmitt, M., B{\"u}rkner, P.-C., K{\"o}the, U., and Radev, S.~T. (2023a).
\newblock Detecting model misspecification in amortized bayesian inference with neural networks.
\newblock In {\em DAGM German Conference on Pattern Recognition}, pages 541--557. Springer.

\bibitem[Schmitt et~al., 2023b]{schmitt_meta-uncertainty_2023}
Schmitt, M., Radev, S.~T., and Bürkner, P.-C. (2023b).
\newblock Meta-{Uncertainty} in {Bayesian} {Model} {Comparison}.
\newblock {\em AISTATS Conference Proceedings}.

\bibitem[Shannon, 1998]{shannon1998introduction}
Shannon, R.~E. (1998).
\newblock Introduction to the art and science of simulation.
\newblock In {\em Winter simulation conference proceedings}, volume~1, pages 7--14. IEEE.

\bibitem[Sharrock et~al., 2022]{sharrock2022diffusionsbi}
Sharrock, L., Simons, J., Liu, S., and Beaumont, M. (2022).
\newblock Sequential neural score estimation: Likelihood-free inference with conditional score based diffusion models.
\newblock {\em arXiv preprint}.

\bibitem[Simon, 1974]{simon1974sciences}
Simon, H.~A. (1974).
\newblock {\em The Sciences of the Artificial}.
\newblock MIT Press.

\bibitem[Song et~al., 2021]{song2020denoising}
Song, J., Meng, C., and Ermon, S. (2021).
\newblock Denoising diffusion implicit models.
\newblock {\em International Conference on Learning Representations}.

\bibitem[Sountsov et~al., 2024]{sountsov2024running}
Sountsov, P., Carroll, C., and Hoffman, M.~D. (2024).
\newblock Running markov chain monte carlo on modern hardware and software.
\newblock {\em arXiv preprint arXiv:2411.04260}.

\bibitem[South et~al., 2023]{south2023regularized}
South, L.~F., Oates, C.~J., Mira, A., and Drovandi, C. (2023).
\newblock Regularized zero-variance control variates.
\newblock {\em Bayesian Analysis}, 18(3):865--888.

\bibitem[Stuhlm{\"u}ller et~al., 2013]{stuhlmuller2013learning}
Stuhlm{\"u}ller, A., Taylor, J., and Goodman, N. (2013).
\newblock Learning stochastic inverses.
\newblock {\em Advances in neural information processing systems}, 26.

\bibitem[Swierc et~al., 2024]{swierc2024domain}
Swierc, P., Tamargo-Arizmendi, M., {\'C}iprijanovi{\'c}, A., and Nord, B.~D. (2024).
\newblock Domain-adaptive neural posterior estimation for strong gravitational lens analysis.
\newblock {\em arXiv preprint arXiv:2410.16347}.

\bibitem[Säilynoja et~al., 2022]{sailynoja_graphical_2022}
Säilynoja, T., Bürkner, P.-C., and Vehtari, A. (2022).
\newblock Graphical test for discrete uniformity and its applications in goodness-of-fit evaluation and multiple sample comparison.
\newblock {\em Statistics and Computing}, 32(2).

\bibitem[Säilynoja et~al., 2025]{sailynoja2025posteriorsbc}
Säilynoja, T., Schmitt, M., Bürkner, P.-C., and Vehtari, A. (2025).
\newblock Posterior sbc: Simulation-based calibration checking conditional on data.
\newblock {\em arXiv preprint:2502.03279}.

\bibitem[Talts et~al., 2018]{talts_validating_2018}
Talts, S., Betancourt, M., Simpson, D., Vehtari, A., and Gelman, A. (2018).
\newblock Validating {Bayesian} inference algorithms with simulation-based calibration.
\newblock {\em arXiv preprint:1804.06788}.

\bibitem[Tavar{\'e} et~al., 1997]{tavare1997inferring}
Tavar{\'e}, S., Balding, D.~J., Griffiths, R.~C., and Donnelly, P. (1997).
\newblock Inferring coalescence times from dna sequence data.
\newblock {\em Genetics}, 145(2):505--518.

\bibitem[Tierney, 1994]{tierney1994markov}
Tierney, L. (1994).
\newblock Markov chains for exploring posterior distributions.
\newblock {\em the Annals of Statistics}, pages 1701--1728.

\bibitem[Tokdar and Kass, 2010]{tokdar2010importance}
Tokdar, S.~T. and Kass, R.~E. (2010).
\newblock Importance sampling: a review.
\newblock {\em Wiley Interdisciplinary Reviews: Computational Statistics}, 2(1):54--60.

\bibitem[Van~Dongen, 2006]{van_dongen_prior_2006}
Van~Dongen, S. (2006).
\newblock Prior specification in {Bayesian} statistics: {Three} cautionary tales.
\newblock {\em Journal of Theoretical Biology}, 242(1):90--100.

\bibitem[Vasishth and Broe, 2010]{vasishth_foundations_2010}
Vasishth, S. and Broe, M. (2010).
\newblock {\em The {Foundations} of {Statistics}: {A} {Simulation}-based {Approach}}.
\newblock Springer Science \& Business Media.

\bibitem[Vehtari et~al., 2017]{vehtari_practical_2017}
Vehtari, A., Gelman, A., and Gabry, J. (2017).
\newblock Practical {Bayesian} model evaluation using leave-one-out cross-validation and {WAIC}.
\newblock {\em Statistics and Computing}, 27(5):1413--1432.

\bibitem[Vehtari et~al., 2024]{vehtari2024pareto}
Vehtari, A., Simpson, D., Gelman, A., Yao, Y., and Gabry, J. (2024).
\newblock Pareto smoothed importance sampling.
\newblock {\em Journal of Machine Learning Research}, 25(72):1--58.

\bibitem[Vehtari et~al., 2019]{vehtari2019limitations}
Vehtari, A., Simpson, D.~P., Yao, Y., and Gelman, A. (2019).
\newblock Limitations of “limitations of bayesian leave-one-out cross-validation for model selection”.
\newblock {\em Computational Brain \& Behavior}, 2(1):22--27.

\bibitem[von Krause et~al., 2022]{von2022mental}
von Krause, M., Radev, S.~T., and Voss, A. (2022).
\newblock Mental speed is high until age 60 as revealed by analysis of over a million participants.
\newblock {\em Nature Human Behaviour}, 6(5):700--708.

\bibitem[Wagenmakers et~al., 2022]{Wagenmakers2022}
Wagenmakers, E.-J., Sarafoglou, A., and Aczel, B. (2022).
\newblock One statistical analysis must not rule them all.
\newblock {\em Nature}, 605(7910):423–425.

\bibitem[Xu et~al., 2020]{xu2020can}
Xu, K., Li, J., Zhang, M., Du, S.~S., Kawarabayashi, K.-i., and Jegelka, S. (2020).
\newblock What can neural networks reason about?
\newblock In {\em International Conference on Learning Representations}.

\bibitem[Yao and Domke, 2023]{yao2023discriminative}
Yao, Y. and Domke, J. (2023).
\newblock Discriminative calibration: Check bayesian computation from simulations and flexible classifier.
\newblock {\em Advances in Neural Information Processing Systems}, 36:36106--36131.

\bibitem[Zammit-Mangion et~al., 2024]{zammit2024neural}
Zammit-Mangion, A., Sainsbury-Dale, M., and Huser, R. (2024).
\newblock Neural methods for amortized inference.
\newblock {\em Annual Review of Statistics and Its Application}, 12.

\bibitem[Zeng et~al., 2025]{zeng2025real}
Zeng, J., Xue, K., and Chen, H. (2025).
\newblock Real-time probabilistic model updating and damage detection using machine learning-based likelihood-free inference.
\newblock {\em Mechanical Systems and Signal Processing}, 230:112612.

\bibitem[Zitzmann et~al., 2021]{zitzmann_prior_2021}
Zitzmann, S., Helm, C., and Hecht, M. (2021).
\newblock Prior {Specification} for {More} {Stable} {Bayesian} {Estimation} of {Multilevel} {Latent} {Variable} {Models} in {Small} {Samples}: {A} {Comparative} {Investigation} of {Two} {Different} {Approaches}.
\newblock {\em Frontiers in Psychology}, 11:611267.

\end{thebibliography}

\end{document}